\documentclass[a4paper,11pt]{article}
\pdfoutput=1
\usepackage{jheppub}

\usepackage{amssymb,amsmath,amsfonts}
\usepackage[normalem]{ulem}
\usepackage[utf8x]{inputenc}
\usepackage{slashed}
\usepackage{graphicx}
\usepackage{tabularx}
\usepackage{here}
\usepackage{color}
\usepackage{csquotes} 
\usepackage{comment}
\usepackage{mathrsfs}
\usepackage{float}
\usepackage{ascmac}
\usepackage{multirow}
\usepackage{longtable}
\usepackage{bm}
\usepackage{ulem}

\usepackage[italicdiff]{physics}

\makeatletter
\newcommand*\rel@kern[1]{\kern#1\dimexpr\macc@kerna}
\newcommand*\widebar[1]{%
  \begingroup
  \def\mathaccent##1##2{%
    \rel@kern{0.8}%
    \overline{\rel@kern{-0.8}\macc@nucleus\rel@kern{0.2}}%
    \rel@kern{-0.2}%
  }%
  \macc@depth\@ne
  \let\math@bgroup\@empty \let\math@egroup\macc@set@skewchar
  \mathsurround\z@ \frozen@everymath{\mathgroup\macc@group\relax}%
  \macc@set@skewchar\relax
  \let\mathaccentV\macc@nested@a
  \macc@nested@a\relax111{#1}%
  \endgroup
}
\makeatother

\numberwithin{equation}{section}

\preprint{
\begin{minipage}{5cm}
\small
\flushright
EPHOU-25-012\\
KYUSHU-HET-329
\end{minipage}}

\title{
On discrete gauging and non-invertible selection rules}

\author{Jun Dong$^{1}$,} 
\author{Tim Jeric$^{1}$,}
\author{Tatsuo Kobayashi$^{1}$,} 
\author{Ryusei Nishida$^{1}$, and } 
\author{Hajime Otsuka$^{2}$}
\affiliation{
$^1$Department of Physics, Hokkaido University, Sapporo 060-0810, Japan\\
$^2$Department of Physics, Kyushu University, 744 Motooka, Nishi-ku, Fukuoka 819-0395, Japan
}
\emailAdd{j-dong@particle.sci.hokudai.ac.jp}
\emailAdd{t-jeric@particle.sci.hokudai.ac.jp}
\emailAdd{kobayashi@particle.sci.hokudai.ac.jp}
\emailAdd{r-nishida@particle.sci.hokudai.ac.jp}
\emailAdd{otsuka.hajime@phys.kyushu-u.ac.jp}

\abstract{
We clarify selection rules of conjugacy classes of several finite discrete groups where we deal with both gauged and ungauged cases. 
We find that the selection rules enjoy finite Abelian or non-Abelian discrete symmetries originating from the inner and/or outer automorphism of underlying discrete groups. 
Since the selection rules of conjugacy classes do not obey conventional group-like selection rules, they open up new coupling selection rules of fields which are labeled by the conjugacy classes. 
}

\makeatletter
\gdef\@fpheader{}
\makeatother

\begin{document}

\maketitle

\section{Introduction}
\label{sec:Intro}

In string theory on orbifolds, some string states can be labeled by conjugacy classes of a group rather than its element due to the identification of string states under the orbifold twist. 
For example, suppose that the string has the boundary condition, 
$X(\sigma = \pi) =gX(\sigma=0)$, where $g$ denotes the space group action constructed by the point group action and a lattice translation. The same boundary condition may be satisfied by $hX$ as 
$hX(\sigma = \pi) =ghX(\sigma=0)$ with $h$ being an additional space group action. That is, the boundary condition represented by $g$ must be equivalent to $h^{-1}gh$, and equivalent elements belong to a single class. Similarly, if compact space and world-sheet theory have a symmetry, we can divide them by such a symmetry. Hence, we can gauge such a symmetry so as to obtain classes of string states and vertex operators.
The selection rules of conjugacy classes do not obey a group-like selection rule but a non-invertible one in general. 
Indeed, such non-invertible selection rules are known in heterotic string theory on toroidal orbifolds~\cite{Dijkgraaf:1987vp,Kobayashi:2004ya, Kobayashi:2006wq,Beye:2014nxa,Thorngren:2021yso,Heckman:2024obe,Kaidi:2024wio} and type II intersecting/magnetized D-brane models~\cite{Kobayashi:2024yqq,Funakoshi:2024uvy} (see, e.g., Refs.~\cite{Verlinde:1988sn,Moore:1988qv,Moore:1989yh,Fuchs:1993et,Bhardwaj:2017xup} for the fusion rules in two-dimensional conformal field theories.). Furthermore, it can also be seen in Calabi-Yau compactifications~\cite{Dong:2025pah}. 
Indeed, a translation is invertible on the $T^2$ torus, but it is not a gauge invariant quantity under an orbifold group~\cite{Heckman:2024obe}. 
Hence, in toroidal orbifolds, the momentum operators are constructed in a way to be invariant under the orbifold twists.
In the type IIB magnetized D-brane models on $T^2/\mathbb{Z}_N$ with magnetic fluxes, the $\mathbb{Z}_N$ invariant momentum operators obey non-invertible fusion rules under which degenerate chiral zero-modes non-trivially transform under the non-invertible symmetries. 
See for reviews of non-invertible symmetries, e.g., Refs.~\cite{Schafer-Nameki:2023jdn,Shao:2023gho}. Orbifolding is a way to obtain non-invertible fusion rules.

When the fields in quantum field theory are labeled by the elements in the algebra obeying the non-invertible selection rules, it exhibits interesting phenomenological applications in particle physics~\cite{Choi:2022jqy,Cordova:2022fhg,Cordova:2022ieu,Cordova:2024ypu,Kobayashi:2024cvp,Kobayashi:2024yqq,Kobayashi:2025znw,Suzuki:2025oov,Liang:2025dkm,Kobayashi:2025ldi,Kobayashi:2025cwx,Nomura:2025sod,Kobayashi:2025lar}. 
For instance, let us suppose that quarks and leptons are labeled by classes obeying non-invertible selection rules in a family-dependent way. Then, it leads to the different structure of the textures of Yukawa matrix from models with conventional flavor symmetries. Indeed, one can realize the realistic fermion masses and mixings as well as CP phases in the quark~\cite{Kobayashi:2024cvp,Kobayashi:2025znw} and lepton sectors~\cite{Kobayashi:2025ldi}. 
Furthermore, one can realize the interesting Yukawa texture of quarks, addressing the strong CP problem without axion in the context of spontaneous CP violation~\cite{Kobayashi:2025znw,Liang:2025dkm}. 
Such non-invertible selection rules can be violated by radiative and/or stringy corrections~\cite{Heckman:2024obe,Kaidi:2024wio,Funakoshi:2024uvy}; thereby certain couplings are generated at loop levels. It will be the reason why the masses of active neutrinos are tiny. It was explicitly checked in the context of radiative neutrino mass models~\cite{Kobayashi:2025cwx,Nomura:2025sod}. 

The purpose of this paper is to clarify the selection rules of conjugacy classes of finite discrete groups and reveal remaining symmetries originating from the inner and/or outer automorphism of underlying discrete groups. 
Concerning the conjugacy classes of a finite discrete group $G$, we also discuss the gauging of $G$. The $\mathbb{Z}_2$ gauging of $\mathbb{Z}_N$ symmetries are discussed in Refs.~\cite{Kobayashi:2024yqq,Kobayashi:2024cvp}, corresponding to $\mathbb{Z}_2$-invariant modes under the orbifold twist. 
It was found that such a $\mathbb{Z}_2$ gauging is phenomenologically useful to realize the interesting Yukawa textures. 
In this paper, we extend this analysis to $\mathbb{Z}_3$ and $S_3$ gaugings of a certain finite discrete group. 
Then, we apply the selection rules obtained from $\mathbb{Z}_3$ gauging of $\mathbb{Z}_N$ symmetry to phenomenological models where fields are labeled by the elements in the algebra. It leads to different properties between $\mathbb{Z}_2$ and $\mathbb{Z}_3$ gauging of $\mathbb{Z}_N$ symmetry. 

This paper is organized as follows. 
In Sec.~\ref{sec:Z2-gauge-DN}, we first review the selection rules of classes for $\mathbb{Z}_2$ gauging of $\mathbb{Z}_N$ symmetries and extend to the $D_N$ conjugacy classes. 
This discussion is further extended to $\mathbb{Z}_3$ gauging of $\mathbb{Z}_N$ symmetries in Sec.~\ref{sec:Z3gauging-0}. 
In Sec.~\ref{sec:Z3gauge}, we discuss the conjugacy classes for $\mathbb{Z}_3$ gauging of $\mathbb{Z}_N\times \mathbb{Z}'_N$ and $\Delta(3N^2)$. 
In Sec.~\ref{sec:Z3NNN}, we deal with the selection rules of the conjugacy classes for $\mathbb{Z}_3$ gauging of $\mathbb{Z}_N\times \mathbb{Z}'_N \times \mathbb{Z}''_N$ and $\Sigma(3N^3)$. 
The selection rules of the conjugacy classes for $S_3$ gauging of $\mathbb{Z}_N\times \mathbb{Z}'_N$ and $\Delta(6N^2)$ are discussed in Sec.~\ref{sec:S3gauge}. 
In Sec.~\ref{sec:pheno},  we discuss particle phenomenological aspects of multiplication rules, which were obtained in the previous sections. 
Finally, Sec.~\ref{sec:con} is devoted to the conclusions. 
In Appendix~\ref{app:Z3-gauge-9}, we list the selection rules of classes for $\mathbb{Z}_3$ gauging of $\mathbb{Z}_9$ symmetry, and $\Delta(12)\cong A_4$ and $\Delta(24)\cong S_4$ cases are summarized in Appendix~\ref{app:A4S4}. 
In Appendices~\ref{app:futher} and~\ref{app:Z2}, we discuss other types of gauging.

\section{$\mathbb{Z}_2$ gauging of $\mathbb{Z}_N$ and $D_N$}
\label{sec:Z2-gauge-DN}

\subsection{$\mathbb{Z}_2$ gauging of $\mathbb{Z}_N$}
\label{sec:Z2gauge}

Here, we review about $\mathbb{Z}_2$ gauging of $\mathbb{Z}_N$ symmetries \cite{Kobayashi:2024yqq,Kobayashi:2024cvp}.
We start with the $\mathbb{Z}_N$ symmetry.
We denote its generator by $a$, which satisfies 
\begin{align}
    a^N=e,
\end{align}
where $e$ is the identity.
All of the $\mathbb{Z}_N$ elements are written by $a^k$ ($k=0,1,\cdots, N-1$) (mod $N$).
Their multiplication rules are written as 
\begin{align}
    a^{k_1}a^{k_2}=a^{k_1+k_2}.
\label{eq:ZNrule}
\end{align}

Next, we explain $\mathbb{Z}_2$ gauging.
We introduce the $\mathbb{Z}_2$ element $b$, which satisfies 
\begin{align}
    b^2=e, \qquad ba^k b^{-1}=a^{-k}.
\end{align}
That is, $b$ corresponds to the outer automorphism of the $\mathbb{Z}_N$ group.
Then, we define the class $C^{(k)}$ as 
\begin{align}
    C^{(k)} = \{ b^n a^kn^{-n} | n=0,1 \}= \{ a^k, a^{-k} \}.
\end{align}
That is $\mathbb{Z}_2$ gauging of the $\mathbb{Z}_N$ symmetry.
These classes satisfy the following fusion rule:
\begin{align}\label{eq:Z2gauge-rule}
    C^{(k)}\cdot C^{(k')}=C^{(k+k')}+C^{(k-k')}.
\end{align}
They are commutable. 
Similarly, multiplication rules, which we will derive in the following sections, are commutable.

When $N=$ even, this algebra has the $\mathbb{Z}_2$ symmetry $C^{(k=\rm odd)}\rightarrow - C^{(k=\rm odd)}$, 
because this leads the following relations:
\begin{align}\label{eq:Z2-inner}
     &C^{(k={\rm even})}\cdot C^{(k={\rm even})}=C^{(k={\rm even})}+C^{(k=\rm even)}, \notag \\ 
         &C^{(k={\rm even})}\cdot C^{(k={\rm odd})}=C^{(k={\rm odd})}+C^{(k=\rm odd)}, \\
          &C^{(k={\rm odd})}\cdot C^{(k={\rm odd})}=C^{(k={\rm even})}+C^{(k=\rm even)}.   \notag 
\end{align}
When $N=$ odd, there is no such a symmetry.

In general, the class $C^{(k)}$ includes two elements, $a^k$ and $a^{-k}$, but obviously for $k=0$, the class $C^{(0)}$ includes only one element $a^0=e$.
In addition, when $N$ is even, the class $C^{(N/2)}$ also includes only one element.
We denote them as 
\begin{align}
    C_1 = \{ e\}, \qquad C_1^{(N/2)}=\{ a^{N/2}\}, \qquad C_2^{(k)}=C^{(k)}, 
\end{align}
for $k \neq 0,N/2$.
They satisfy
\begin{align}
    &C_1\cdot C_1=C_1^{(N/2)}\cdot C_1^{(N/2)}=C_1, \notag \\ 
    &C_1\cdot C_2^{(k)}=C_1^{(N/2)}\cdot C_2^{(k)}=C_2^{(k)}, \\
      &C_2^{(k)}\cdot C_2^{(k')}=C_2^{(k+k')}+C_2^{(k-k')}, \notag
\end{align}
where $k+k'\neq 0$ and $k-k'\neq 0$ (mod $N/2$).
When $k+k'= 0$ and $k-k'= 0$ (mod $N$),  
$C_2^{(k+k')}$ and $C_2^{(k-k')}$ in the right hand side of the last equation must be replaced by $2C_1$.
Similarly, when $k+k'= N/2$ and $k-k'= N/2$ (mod $N$), $C_2^{(k+k')}$ and $C_2^{(k-k')}$ in the right hand side of the last equation must be replaced by $2C_1^{(N/2)}$.
$\mathbb{Z}_2$ gauging of $\mathbb{Z}_N$ symmetries can be realized by string compactifications \cite{Kobayashi:2024yqq,Funakoshi:2024uvy}.
Certain compactifications such as  magnetized compactifications of higher-dimensional Yang-Mills theory and string theory lead to the 
$\mathbb{Z}_N$ symmetry \cite{Abe:2009vi,Berasaluce-Gonzalez:2012abm,Marchesano:2013ega}.
In these models, the massless modes $\varphi_k$ transform as 
\begin{align}\label{eq:ZN-mode}
    \varphi_k \to a^k \varphi_k,
\end{align}
under the $\mathbb{Z}_N$ symmetry.
The $\mathbb{Z}_N$ charge $k$ corresponds to discrete momentum, winding number, and their mixture.

Now, we consider geometrical $\mathbb{Z}_2$ orbifolding of the above compactification.
Then, the $\mathbb{Z}_2$-invariant modes are written by \cite{Abe:2008fi}
\begin{align}
\label{eq:Z2-inv-field}
    \phi_k = \varphi_k + \varphi_{N-k},
\end{align}
up to normalization.
These modes $\phi_k$ correspond to $C^{(k)}$.
On this $\mathbb{Z}_2$ orbifold compactification, the original $\mathbb{Z}_N$ symmetry is broken, but 
certain coupling selection rules remain in low energy effective field theory.
That corresponds to Eq.~(\ref{eq:Z2gauge-rule}).
The mode $\phi_k$ behaves as if it has both $k$ and $N-k$ charges under $\mathbb{Z}_N$.

Each field $\phi_k$ corresponds to a class $C^{(k)}$.
Note that the field $\phi_k$ and its conjugate $\phi_k^*$ correspond to the same class $C^{(k)}$.
Their coupling selection rules are determined by fusion rules (\ref{eq:Z2gauge-rule}).
For example, we consider the two-point coupling of $\phi_k$ and $\phi_{k'}$, which correspond to the classes $C^{(k)}$ and $C^{(k')}$.
If their multiplication $C^{(k)}\cdot C^{(k')}$ includes the identity $e$, their two-point coupling is allowed.
The condition is written by
\begin{align}
    \pm k \pm k'=0 ~~({\rm mod}~~N).
\end{align}
We find that $C^{(k)}\cdot C^{(k)}$ always include $e$ and the 
two-point coupling $\phi_k \phi_k$ such as mass terms and kinetic terms are allowed.
That leads the property that the diagonal entries of neutrino mass matrix induced by the Weinberg operators are always allowed \cite{Kobayashi:2025ldi}.
Also, $2n$ point self couplings $(\phi_k)^{2n}$ are allowed.

Similarly, we can study the $n$-point couplings, $\phi_{k_1}\cdots \phi_{k_n}$.
Their couplings are allowed if $C^{(k_1)}\cdots C^{(k_n)}$ includes the identity.
The condition can be written by
\begin{align}
    \sum_i \pm k_i=0 ~~({\rm mod}~~N).
\end{align}
For example, when $N=4$,  
there are three classes: 
\begin{align}\label{eq:class-Z2-4}
  &C_1 = \{ e \},\quad
  C_2^{(1)} = \{ a,a^3 \},\quad
  C_1^{(2)} = \{ a^2 \},
\end{align}
and they obey the fusion rules as presented in Table~\ref{tab:Z2-Z2}. 
It is nothing but the Ising fusion rule realized in $c=\frac{1}{2}$ critical Ising model with three primary operators $\{\mathbb{I}, \epsilon, \sigma\}$:
\begin{align}
    \epsilon \cdot \epsilon = \mathbb{I}\,,
\quad
    \epsilon \cdot \sigma = \sigma \cdot \epsilon = \sigma\,,
\quad
    \sigma \cdot \sigma = \mathbb{I} + \epsilon\,,
\end{align}
corresponding to $C_1= \mathbb{I}$, $C_1^{(2)}=\epsilon$ and $C_2^{(1)}=\sqrt{2}\sigma$. 
As mentioned in Eq.~(\ref{eq:Z2-inner}), their multiplication rules have the $\mathbb{Z}_2$ symmetry, where $C_2^{(1)}$ is $\mathbb{Z}_2$ odd, while $C_1$ and $C_1^{(2)}$ are $\mathbb{Z}_2$ even.\footnote{Similarly, one can realize the Fibonacci fusion rule for $N=3$ and the fusion rule of $SU(2)_3$ for $N=6$.} 

\begin{table}[H]
    \centering
    \caption{Multiplication rules for $\mathbb{Z}_2$ gauging of  $\mathbb{Z}_4$.}
    \label{tab:Z2-Z2}
    \begin{tabular}{|c||c|c|c|}
    \hline
         &  $C_1$ & $C^{(1)}_{2}$ & $C^{(2)}_1$  \\
         \hline\hline
         $C_1$&  $ C_1 $& $ C^{(1)}_2 $ & $C^{(2)}_1  $   \\
         \hline
         $C^{(1)}_{2}$&  $C^{(1)}_{2}$  & $ 2C_1+2C^{(2)}_1 $ & $ C^{(1)}_2 $  \\
         \hline
         $C^{(2)}_1$&  $C^{(2)}_1$ & $ C^{(1)}_2 $ & $ C_1 $ \\
         \hline
    \end{tabular}
\end{table}

It is very important to apply the above section rules due to $\mathbb{Z}_2$ gauging of $\mathbb{Z}_N$ symmetries to particle phenomenology.
For example, we can derive interesting texture structures by $\mathbb{Z}_2$ gauging of $\mathbb{Z}_N$ symmetries \cite{Kobayashi:2024cvp,Kobayashi:2025znw,Liang:2025dkm,Kobayashi:2025ldi}.
We can also study radiative neutrino mass models \cite{Kobayashi:2025cwx} and further applications.
In the following sections, we will extend $\mathbb{Z}_2$ gauging of $\mathbb{Z}_N$ symmetries to $\mathbb{Z}_3$ gauging of discrete symmetries as well as $S_3$ gauging.

\subsection{$D_N$}

Before extending $\mathbb{Z}_2$ gauging to $\mathbb{Z}_3$ gauging, we discuss the relations of $\mathbb{Z}_2$ gauging of the $\mathbb{Z}_N$ symmetries to the $D_N=\mathbb{Z}_N \rtimes \mathbb{Z}_2$ symmetries.
The classes, $C_1$, $C_1^{(N/2)}$, and $C_2^{(k)}$ in the previous section correspond to the conjugacy classes of $D_N$, which do not include $b$ \cite{Ishimori:2010au,Kobayashi:2022moq}.\footnote{Here, we follow the notation in Refs.~\cite{Ishimori:2010au,Kobayashi:2022moq}. In some references, $D_N$ is written as $D_{2N}$.}
They also correspond to the $\mathbb{Z}_2$-invariant modes (\ref{eq:Z2-inv-field}).

The $D_N$ group has other conjugacy classes.
When $N=$odd, there is another conjugacy class:
\begin{align}
    B_1\equiv \{b, ab,ab^2, ..., ba^{N-1} \}.
\end{align}
There are trivial and non-trivial singlets and doublets.

When $N=$even, there are two conjugacy classes including $b$, 
\begin{align}
    B_1\equiv \{b, a^2b, ..., a^{N-2}b \},
    \qquad
    B_2\equiv \{ab, a^3b, ..., a^{N-1}b \},
\end{align}
in addition to $C^{(k)}$. 
There are four singlets in addition to doublets.

For $C^{(k)}$, we have the realization by using the "bulk" in Eq.~(\ref{eq:Z2-inv-field}).
We may need to introduce the twist fields, which change  the boundary condition from $\mathbb{Z}_2$ even to $\mathbb{Z}_2$ odd \cite{Hamidi:1986vh,Dixon:1986qv}.
For example, when $N=4$, there are two conjugacy classes including $b$, $B_1=\{b, a^2b \}$ and $B_2=\{ ab \}$.
Then, we introduce two twist fields, $\sigma_0$ and $\sigma_1$, 
which correspond to $a^kb$ with $k=$even and odd, respectively, where $k$ corresponds to the momentum-winding number.
This structure is consistent with the world-sheet theory on $S^1/\mathbb{Z}_2$.
The fixed points on $S^1/\mathbb{Z}_2$ are represented by 
the conjugacy class of the space group $(\theta,ne)$, where 
$\theta$ denotes $\mathbb{Z}_2$ twist and $e$ denotes the lattice basis vector to construct $S^1$.
There are two conjugacy classes $(\theta,ne)$ with $n=0,1$.
They correspond to $\sigma_n$ with $n=0,1$, and $B_1$ and $B_2$.
There is the $\mathbb{Z}_2$ symmetry related to the generator $a$.
Under this $\mathbb{Z}_2$, the twist fields transform as
\begin{align}\label{eq:twist-1}
    \sigma_n \to (-1)^n \sigma_n.
\end{align}
Also there is the following permutation symmetry:
\begin{align}\label{eq:twist-2}
    \sigma_0 \to \sigma_1, \qquad \sigma_1 \to \sigma_0,
\end{align}
because the geometry does not change under the exchange of fixed points.
Then, the full symmetry becomes the $D_4$ symmetry \cite{Dijkgraaf:1987vp,Thorngren:2021yso,Heckman:2024obe,Kaidi:2024wio,Kobayashi:2004ya,Kobayashi:2006wq,Beye:2014nxa}.

\begin{table}[H]
    \centering
    \caption{Multiplication rules for conjugacy classes of $D_4$.}
    \label{tab:D4}
    \begin{tabular}{|c||c|c|c|c|c|}
    \hline
         &  $C_1$ & $C^{(1)}_{2}$ & $C^{(2)}_1$ & $B_1$ & $B_2$ \\
         \hline\hline
         $C_1$&  $ C_1 $& $ C^{(1)}_2 $ & $C^{(2)}_1  $ & $ B_1 $ & $ B_2 $   \\
         \hline
         $C^{(1)}_{2}$&  $C^{(1)}_{2}$  & $ 2C_1+2C^{(2)}_1 $ & $ C^{(1)}_2 $ & $ B_1 $ &  $ B_2 $  \\
         \hline
         $C^{(2)}_1$&  $C^{(2)}_1$ & $ C^{(1)}_2 $ & $ C_1 $ & $ B_1 $ &  $ B_2 $ \\
         \hline
         $B_1$&  $B_1$  & $ B_1 $ & $ B_1 $ & $ 2C_1+2C^{(2)}_1 $ & $2C^{(1)}_2$  \\
         \hline
         $B_2$&  $B_2$  & $ B_2 $ & $ B_2 $ & $2C^{(1)}_2  $ & $ 2C_1+2C^{(2)}_1 $ \\
         \hline
    \end{tabular}
\end{table}

The multiplication rules of $D_4$ conjugacy classes have several symmetries. 
One is the permutation symmetry $B_1 \leftrightarrow B_2$, which corresponds to Eq.~(\ref{eq:twist-2}). 
In addition, recall that when $N=$ even, there is the $\mathbb{Z}_2$ symmetry, where $C^{(k={\rm even})}$ and $C^{(k={\rm odd})}$ respectively have $\mathbb{Z}_2$ even and odd charges. To be consistent, $B_1$ and $B_2$ must have $\mathbb{Z}_2$ even and odd charges , respectively. 
That corresponds to Eq.~(\ref{eq:twist-1}). 
We combine them as to find that the full symmetry is $D_4$.
The conjugacy classes $B_1$ and $B_2$ are $D_4$ doublets, while $C_2^{(1)}$ is non-trivial singlet and $C_1$ and $C_1^{(2)}$ are trivial singlets. 
On the other hand, when $N=$ odd, we may introduce a single twist field $\sigma$ in order to represent the conjugacy class $B_1$.
It may correspond to one of $\sigma_0$ and $\sigma_1$, or both, but its geometrical meaning is not clear.

At any rate, $\mathbb{Z}_2$ invariant elements correspond to $C^{(k)}$ and they satisfy the algebraic relations (\ref{eq:Z2gauge-rule}).

\section{$\mathbb{Z}_3$ gauging of $\mathbb{Z}_N$}
\label{sec:Z3gauging-0}

We study the extension of $\mathbb{Z}_2$ gauging in the previous section, e.g. to $\mathbb{Z}_3$ gauging.
The simplest extension is $\mathbb{Z}_3$ gauging of $\mathbb{Z}_N$ symmetries, whose generator is $a$, i.e. $a^N=e$.
We consider the $\mathbb{Z}_3$ automorphism $b$ of the $\mathbb{Z}_N$ group, i.e.,
\begin{align}\label{eq:Z3-ZN}
      b^{-1}ab=a^m, \qquad b^3=e,
\end{align}
for $m \neq 0$.
Since the case with $m=1$ $({\rm mod}~N)$ is trivial, we study other numbers.
Eq.~(\ref{eq:Z3-ZN}) can be written as 
\begin{align}
    a=ba^mb^{-1}.
\end{align}
By use of this, we can derive
\begin{align}
    b^{-1}ab=a^m=(ba^mb^{-1})^m=(ba^mb^{-1}) \cdots (ba^mb^{-1})=ba^{m^2}b^{-1}.
\end{align}
This relation can be written as
\begin{align}
    a=b^{-1}a^{m^2}b,
\end{align}
which leads to  
\begin{align}
    b^{-1}ab=a^{m}=(b^{-1}a^{m^2}b)^m=(b^{-1}a^{m^2}b) \cdots (b^{-1}a^{m^2}b)=b^{-1}a^{m^3}b.
\end{align}
Through these consistency conditions, it is found that 
\begin{align}
    a^{m^3}=a.
\end{align}
That is, $m$ must satisfy 
\begin{align}\label{eq:condition-Z3}
    m^3-1=(m-1)(m^2+m+1)=0\,\,\,(\mathrm{mod}~N).
\end{align}
One of the solution is $m=1$ (mod $N$), but that is trivial as mentioned above. 
Suppose that $N$ is a prime number. Hence, $m$ must satisfy 
\begin{align}
    m^2+m+1=0\,\,\,(\mathrm{mod}~N).
\end{align}
The possible combinations of $(N,m)$ are 
\begin{align}
    (N,m)=(7,2), (13,3), \cdots.
\end{align}

For generic element $a^k$, we find 
\begin{align}
    &b a^k b^{-1} = a^{m^2 k}, \qquad 
    b^2 a^k b^{-2} = a^{m^4 k}.
\end{align}
Then, we define the following classes:
\begin{align}   \label{eq:class-Z3} 
    &C^{(k)} = [a^k] = \{a^{m^l k} | l = 0, 2, 4\} = \{a^k, a^{km}, a^{km^2}=a^{-(m + 1) k}\}.
\end{align}
Their multiplication rules are written by 
\begin{align}\label{eq:product-Z3}
    C^{(k)}\cdot  C^{(\ell)} = C^{(k + \ell)} + C^{(k + m^2 \ell)} + C^{(k + m^4 \ell)} = C^{(k + \ell)} + C^{(k - (m + 1) \ell)} + C^{(k + m \ell)}.
\end{align}

When $k=0$, the class $C^{(0)}$ includes the only identity.
Thus, we rewrite the classes as 
\begin{align}
    &C_1 = \{ e \},\\
    &C^k_3 = \{ a^k, a^{km}, a^{km^2} \}.
\end{align}
Their multiplication rules can be written by 
\begin{align} 
\label{eq:product-Z3gauging_1}
    &C_1\cdot C^{k}_3 = C^{k}_3,\\
\label{eq:product-Z3gauging_2}
    &C^{k}_3\cdot C^{\ell}_3 = C^{k+\ell}_3+C^{k+\ell m}_3+C^{k+\ell m^2}_3.
\end{align}

For example, when $N=7$ and $m=2$,  
there are three classes: 
\begin{align}\label{eq:class-Z3-7}
  &C_1 = \{ e \},\quad
  C^1_3 = \{ a,a^2,a^4 \},\quad
  C^2_3 = \{ a^3,a^5,a^6 \},
\end{align}
and they obey the fusion rules as presented in Table~\ref{tab:Z3gaugingZ7}. 
It indicates that there exists the $S_2$ permutation symmetry associated with $C_3^1 \leftrightarrow C_3^2$. 
The $S_2$ symmetry is isomorphic to $\mathbb{Z}_2$ symmetry.
In order to emphasize the permutation of classes, we denote $S_2$. 
This $S_2 \simeq \mathbb{Z}_2$ corresponds to the outer automorphism exchanging $a$ and $a^{-1}$, which is used in $\mathbb{Z}_2$ gauging of $\mathbb{Z}_N$ symmetries in the previous section.
\begin{table}[H]
  \centering
  \caption{Multiplication rules of classes for $\mathbb{Z}_3$ gauging of $\mathbb{Z}_7$.}
    \label{tab:Z3gaugingZ7}
  \begin{tabular}{|c||c|c|c|}
  \hline
     & $C_1$ & $C^1_3$ & $C^2_3$ \\
     \hline\hline
     $C_1$&$C_1$ & $C^1_3$ & $C^2_3$ \\
     \hline
     $C^1_3$&$C^1_3$&$C^1_3+2C^2_3$   &$3C_1+C^1_3+C^2_3$  \\
     \hline
     $C^2_3$&$C^2_3$  &$3C_1+C^1_3+C^2_3$  &$2C^1_3+C^2_3$  \\
     \hline
  \end{tabular}
\end{table}

$\mathbb{Z}_3$ gauging of $\mathbb{Z}_N$ can be realized by higher dimensional field theory in a way similar to $\mathbb{Z}_2$ gauging of $\mathbb{Z}_N$.
We start with modes with definite $\mathbb{Z}_N$ charges as Eq.~(\ref{eq:ZN-mode}).
Then, we introduce a proper $\mathbb{Z}_3$ identification, e.g.
\begin{align}
    \phi_1 = \varphi_1 + \varphi_m + \varphi_{m^2}.
\end{align}

The classes in Eq.~(\ref{eq:class-Z3-7}) correspond to 
the conjugacy classes of 
$\mathbb{Z}_7 \rtimes \mathbb{Z}_3=T_7$ without $b$.\footnote{Note that $T_7$ is only simple subgroup of $SU(3)$ having a complex three-dimensional irreducible representation.} 
The full conjugacy classes of $T_7$ include additional classes:
\begin{align}
    C^{(1)}_7=\{ b, ba, \cdots, ba^6 \}, \qquad 
    C^{(2)}_7=\{ b^2, b^2 a, \cdots, b^2 a^6 \}.
\end{align}
Table \ref{tab:T_7} shows multiplication rules of $T_7$ conjugacy classes, which have a $S_2$ symmetry associated with $C_3^1 \leftrightarrow C_3^2$ and $C_7^{(1)} \leftrightarrow C_7^{(2)}$ at the same time. This corresponds to the outer automorphism exchanging $a$ and $a^{-1}$ as mentioned above. 
At any rate, the classes in Eq.~(\ref{eq:class-Z3-7}) correspond to 
$\mathbb{Z}_3$ invariant ones.

\begin{table}[H]
  \centering
  \caption{Multiplication rules for conjugacy classes of $T_7$.}
    \label{tab:T_7}
  \resizebox{\textwidth}{!}{
  \begin{tabular}{|c||c|c|c|c|c|}
  \hline
     & $C_1$ & $C^1_3$ & $C^2_3$& $C^{(1)}_7$&$C^{(2)}_7$ \\
     \hline\hline
     $C_1$&$C_1$ & $C^1_3$ & $C^2_3$&$C^{(1)}_7$ &$C^{(2)}_7$ \\
     \hline
     $C^1_3$&$C^1_3$&$C^1_3+2C^2_3$   &$3C_1+C^1_3+C^2_3$&$3C^{(1)}_7$ &$3C^{(2)}_7$  \\
     \hline
     $C^2_3$&$C^2_3$  &$3C_1+C^1_3+C^2_3$  &$2C^1_3+C^2_3$& $3C^{(1)}_7$&$3C^{(2)}_7$  \\
     \hline
     $C^{(1)}_7$&$C^{(1)}_7$&$3C^{(1)}_7$&$3C^{(2)}_7$&$7C^{(2)}_7$&$7C_1+7C^{(1)}_7+7C^{(2)}_7$\\
     \hline
$C^{(2)}_7$&$C^{(2)}_7$&$3C^{(1)}_7$&$3C^{(2)}_7$&$7C_1+7C^{(1)}_7+7C^{(2)}_7$&$7C^{(1)}_7$\\
     \hline
  \end{tabular}
  }
\end{table}

Each field $\phi_{k}$ can correspond to a class $C^{(k)}$ of $\mathbb{Z}_3$ gauging of $\mathbb{Z}_N$ symmetry.
In $\mathbb{Z}_2$ gauging of $\mathbb{Z}_N$, the field $\phi_{k}$ and its conjugate $\phi_{k}^*$ correspond to the same class, but in $\mathbb{Z}_3$ gauging they are complex conjugate each other. Thus, we can introduce complex representations in a theory with $\mathbb{Z}_3$ gauging of $\mathbb{Z}_N $.
These aspects are different from $\mathbb{Z}_2$ gauging of $\mathbb{Z}_N$.

The coupling selection rules can be derived in a way similar to coupling selection rules due to $\mathbb{Z}_2$ gauging to $\mathbb{Z}_N$ symmetries.
That is, if the products of $\prod_i C^{(k_i)}$ include the identity $e$, the $n$-point coupling $\prod_i \phi_{k_i}$ is allowed.
For example, the two-point coupling $\phi_{k} \phi_{k}$ is not always allowed, because $C^{(k)}\cdot C^{(k)}$ does not include the identity $e$ for generic $k$.
That is different from the coupling selection rules due to $\mathbb{Z}_2$ of $\mathbb{Z}_N$ symmetries.
On the other hand, we find that  $(C^{(k)})^3$ always includes the identity.
Thus, the three-point self coupling $(\phi_{k})^3$ is always allowed.

When we consider the case with non-prime $N$, more combinations of $(N,m)$ are possible,\footnote{For example, $\mathbb{Z}_3$ gauging of $\mathbb{Z}_9$ is possible. See Appendix \ref{app:Z3-gauge-9}.} but still 
$\mathbb{Z}_3$ gauging of $\mathbb{Z}_N$ is possible for limited $N$. 
Thus, we extend $\mathbb{Z}_N$ to $\mathbb{Z}_N \times \mathbb{Z}'_N$ as well as $\mathbb{Z}_N \times \mathbb{Z}'_N \times \mathbb{Z}''_N$ in the following sections.
Their $\mathbb{Z}_3$ gauging is possible for generic $N$.

\section{$\mathbb{Z}_3$ gauging of $\mathbb{Z}_N\times \mathbb{Z}'_N$ and $\Delta(3N^2)$}
\label{sec:Z3gauge}

\subsection{$\mathbb{Z}_3$ gauging of $\mathbb{Z}_N\times \mathbb{Z}'_N$}
\label{subsec:Z3gauge}

In this section, we extend the analysis of Sec. \ref{sec:Z2gauge} to $\mathbb{Z}_3$ gauging of $\mathbb{Z}_N\times \mathbb{Z}'_N$. 
The generators of $\mathbb{Z}_N\times \mathbb{Z}'_N$ are represented by $a$ and $a'$ which satisfy
\begin{align}
    &a^N = a'^N= e,\quad aa' = a'a,
\end{align}
where $e$ is the identity. All of the $\mathbb{Z}_N\times \mathbb{Z}'_N$ elements are written by $a^\ell a'^m$ $(\ell,m=0,1,\cdots,N-1)$ (mod $N$), and they obey the multiplication rules as in Eq.~\eqref{eq:ZNrule}. 

To describe $\mathbb{Z}_3$ gauging, let us introduce the $\mathbb{Z}_3$ element $b$ satisfying
\begin{align}
    &b^3 = e,\quad bab^{-1} = a^{-1}a'^{-1},\quad ba'b^{-1} = a,\nonumber\\
    &b (a^\ell a'^m) b^{-1} = a^{-\ell + m} a'^{-\ell},\quad
    b^2 (a^\ell a'^m) b^{-2} = a^{-m} a'^{\ell - m}.    
\end{align}
It indicates that $b$ corresponds to the outer automorphism of the $\mathbb{Z}_N \times \mathbb{Z}'_N$ group. 
Then, we define the $\mathbb{Z}_3$ invariant class as follows:
  \begin{align}
\label{eq:Z3classofZNZN}
    &C^{(k,\ell)} = \{ b^n a^k a'^\ell b^{-n} | n=0,1,2  \} = 
    \{ a^ka'^\ell, a^{-k+\ell}a'^{-k}, a^{-\ell}a'^{k-\ell} \}.
  \end{align}
In this $\mathbb{Z}_3$ gauging of the  $\mathbb{Z}_N \times \mathbb{Z}'_N$ symmetry, the classes obey the following fusion rule:
  \begin{align}
    &C^{(k,\ell)}\cdot  C^{(m,n)} = C^{(k+m,\ell+n)}+C^{(k-m+n,\ell-m)}+C^{(k-n,\ell+m-n)}.
  \end{align}
In general, the class $C^{(k,\ell)}$ includes three elements, as shown in \eqref{eq:Z3classofZNZN}, but obviously for $k=\ell=0$, the class $C^{(0,0)}$ includes only one element $a^0a'^0=e$. 
In the following, we consider two cases depending on the value of $N$:

\begin{itemize}
  \item $N/3\neq \ $integer 
    
When $N/3$ is not an integer, we have two types of classes: 
  \begin{align}
    &C_1 = \{ e \},\\
    &C^{(k,\ell)}_3 = \{ a^ka'^\ell, a^{-k+\ell}a'^{-k}, a^{-\ell}a'^{k-\ell} \},
  \end{align}
  with $k,l\neq 0$, and they obey
  \begin{align}
    &C_1\cdot C_1 = C_1,\\
    &C_1\cdot C^{(k,\ell)}_3 = C^{(k,\ell)}_3,\\
    &C^{(k,\ell)}_3\cdot C^{(m,n)}_3 = C^{(k+m,\ell+n)}_3+C^{(k-m+n,\ell-m)}_3+C^{(k-n,\ell+m-n)}_3,
  \end{align}
  with $(k,\ell)\neq(0,0),(m,n)\neq(0,0)$. 
  Note that if any ones of $(k+m,\ell+n)$, $(k-m+n,\ell-m)$ and $(k-n,\ell+m-n)$ are equal to $(0,0)$, $C^{(k+m,-\ell+n)}_3$, $C^{(k-m+n,\ell-m)}_3$, and $C^{(k-n,\ell+m-n)}_3$ in the right hand side of the last equation must be replaced by $3C_1$, i.e.,
  $C^{(0,0)}_3 \Rightarrow 3C_1$. 

  \item $N/3= \ $integer 
  
  On the other hand, when $N/3$ is an integer, we have three types of classes:
  \begin{align}
    \begin{array}{ll}
    C_1 = \{ e \},&\\
    C^{s}_1 = \{ a^sa'^{-s} \}& (s = N/2,2N/3),\\
    C^{(k,\ell)}_3 = \{ a^ka'^\ell, a^{-k+\ell}a'^{-k}, a^{-\ell}a'^{k-\ell} \}\,\,& (k,\ell)\neq (N/3,2N/3),(2N/3,N/3),
  \end{array}
  \end{align}
  and they obey
    \begin{align}
    &C_1\cdot C_1 = C_1,\\
    &C_1\cdot C^{s}_1 = C^{s}_1,\\
    &C_1\cdot C^{(k,\ell)}_3 = C^{(k,\ell)}_3,\\
    &C^s_1\cdot C^{s'}_1 = 
  \left\{ \,
    \begin{aligned}
    & C^{2s}_1\quad (s=s') \\
    & C_1\quad (s\neq s') \\
    \end{aligned}
\right.,\\
    &C^s_1\cdot C^{(k,\ell)}_3 = C^{(s+k,-s+\ell)}_3,\\
    &C^{(k,\ell)}_3\cdot C^{(m,n)}_3 = C^{(k+m,\ell+n)}_3+C^{(k-m+n,\ell-m)}_3+C^{(k-n,\ell+m-n)}_3,
  \end{align}
  with $(k,\ell)\neq(0,0),(m,n)\neq(0,0)$. When $(k+m,\ell+n)$,$(k-m+n,\ell-m)$,$(k-n,\ell+m-n)$ are equal to $(0,0)$, $(N/3,2N/3)$ or $(2N/3,N/3)$, $C_3^{(k,\ell)}$ in the right hand side of the last equation is respectively replaced by $3C_1$, $3C^{N/3}_1$ or $3C^{2N/3}_1$, i.e.,
  \begin{align}
    &C^{(0,0)}_3 \Rightarrow 3C_1,
    \quad 
    C^{(N/3,2N/3)}_3 \Rightarrow 3C^{N/3}_1,
    \quad
    C^{(2N/3,N/3)}_3 \Rightarrow 3C^{2N/3}_1.
  \end{align}
\end{itemize}

When $N/3 = {\rm integer}$, this fusion algebra has $\mathbb{Z}_3$ symmetry, because if $k+\ell = \alpha\pmod 3$ and $m+n = \beta\pmod 3$, then $k+\ell+m+n,,k+\ell-2m+n$ and $k+\ell+m-2n$ are all equivalent to  $\alpha+\beta\pmod 3$, i.e.,
\begin{align}\label{eq:mod-3}
    C^{\alpha}\cdot C^{\beta} = C^{\alpha+\beta}\pmod 3.
\end{align}

For concreteness, we present the multiplication rules of the classes for cases with $N=2$ in Table~\ref{tab:Z3gaugingZ2Z2}, $N=3$ in Table~\ref{tab:Z3gaugingZ3Z3} and $N=4$ in Table~\ref{tab:Z3gaugingZ4Z4}. 
The multiplication rules for $N=3$ are equivalent to those of $\mathbb{Z}_3$ gauging of $\mathbb{Z}_9$ shown in Appendix \ref{app:Z3-gauge-9}. It turns out that there exist two $S_2$ symmetries associated with $C_1^1\leftrightarrow C_1^2$ and $C_3^{(1,0)}\leftrightarrow C_3^{(2,0)}$ as seen in Table~\ref{tab:Z3gaugingZ3Z3}, and two $S_2$ symmetries associated with $C_3^{(1,2)} \leftrightarrow C_3^{(2,1)}$ and $C_3^{(1,0)}\leftrightarrow C_3^{(3,0)}$ as seen in Table~\ref{tab:Z3gaugingZ4Z4}. 
They correspond to the outer automorphisms.
One of them exchanges $a \leftrightarrow a^{-1}$ and $a' \leftrightarrow a'^{-1}$, and the other corresponds exchanges $a \leftrightarrow a'$. 
In addition, recall that when $N/3=$ integer, there is the $\mathbb{Z}_3$ symmetry in Eq.~(\ref{eq:mod-3}), where $C_3^{(1,0)}$ and $C_3^{(2,0)}$ transform as 
\begin{align}\label{eq:Z3-C}
   C_3^{(p,0)} \to \omega^p C_3^{(p,0)},
\end{align}
with $\omega=e^{2\pi i/3}$. When we combine the above symmetry with one of  $S_2$ permutation symmetries, i.e., $a \leftrightarrow a^{-1}$ and $a' \leftrightarrow a'^{-1}$, the symmetry is enhanced to $S_3 \simeq \mathbb{Z}_3 \rtimes S_2$. 
The full symmetry is $S_3 \times S_2$.
\begin{itemize}
  \item $N=2$
  \begin{table}[H]
    \centering
    \caption{Multiplication rules of classes for $\mathbb{Z}_3$ gauging of $\mathbb{Z}_2\times\mathbb{Z}'_2$.}
    \label{tab:Z3gaugingZ2Z2}
    \begin{tabular}{|c||c|c|}
    \hline
       & $C_1$ & $C^{(1,0)}_3$ \\
       \hline
       \hline
       $C_1$ & $C_1$ & $C^{(1,0)}_3$ \\
       \hline
       $C^{(1,0)}_3$ &$C^{(1,0)}_3$ &$3C_1+2C^{(1,0)}_3$\\
       \hline
       \end{tabular}
  \end{table}
  \item $N=3$
  \begin{table}[H]
    \centering
    \caption{Multiplication rules of classes for $\mathbb{Z}_3$ gauging of $\mathbb{Z}_3\times\mathbb{Z}'_3$.}
    \label{tab:Z3gaugingZ3Z3}
    \begin{tabular}{|c||c|c|c|c|c|}
    \hline
       & $C_1$ & $C^1_1$ & $C^2_1$ & $C^{(1,0)}_3$ & $C^{(2,0)}_3$ \\
       \hline
       \hline
       $C_1$& $C_1$ & $C^1_1$ & $C^2_1$ & $C^{(1,0)}_3$ & $C^{(2,0)}_3$ \\
       \hline
       $C^1_1$& $C^1_1$& $C^2_1$ & $C_1$ & $C^{(1,0)}_3$ & $C^{(2,0)}_3$ \\
       \hline
       $C^2_1$& $C^2_1$ &$C_1$  & $C^1_1$ & $C^{(1,0)}_3$ & $C^{(2,0)}_3$ \\
       \hline
       $C^{(1,0)}_3$& $C^{(1,0)}_3$ & $C^{(1,0)}_3$ & $C^{(1,0)}_3$ & $3C^{(2,0)}_3$ & $3C_1+3C^1_1+3C^2_1$ \\
       \hline
       $C^{(2,0)}_3$& $C^{(2,0)}_3$ & $C^{(2,0)}_3$ &$C^{(2,0)}_3$  & $3C_1+3C^1_1+3C^2_1$ & $3C^{(1,0)}_3$ \\
       \hline  
       \end{tabular}
  \end{table}
  \item $N=4$
  \begin{table}[H]
    \centering
    \caption{Multiplication rules of classes for $\mathbb{Z}_3$ gauging of $\mathbb{Z}_4\times\mathbb{Z}'_4$.}
    \label{tab:Z3gaugingZ4Z4}
    \resizebox{\textwidth}{!}{
    \begin{tabular}{|c||c|c|c|c|c|c|}
    \hline
       & $C_1$ & $C^{(1,0)}_3$ & $C^{(2,0)}_3$ & $C^{(3,0)}_3$ & $C^{(1,2)}_3$ & $C^{(2,1)}_3$ \\
       \hline
       \hline
       $C_1$& $C_1$ &$C^{(1,0)}_3$ & $C^{(2,0)}_3$ & $C^{(3,0)}_3$ & $C^{(1,2)}_3$ & $C^{(2,1)}_3$  \\
       \hline
       $C^{(1,0)}_3$& $C^{(1,0)}_3$ & $C^{(2,0)}_3+2C^{(3,0)}_3$ & $C^{(3,0)}_3+C^{(1,2)}_3+C^{(2,1)}_3$ & $3C_1+C^{(1,2)}_3+C^{(2,1)}_3$ & $C^{(1,0)}_3+C^{(2,0)}_3+C^{(1,2)}_3$ & $C^{(1,0)}_3+C^{(2,0)}_3+C^{(2,1)}_3$ \\
       \hline
       $C^{(2,0)}_3$& $C^{(2,0)}_3$ & $C^{(3,0)}_3+C^{(1,2)}_3+C^{(2,1)}_3$ & $3C_1+2C^{(2,0)}_3$ & $C^{(1,0)}_3+C^{(1,2)}_3+C^{(2,1)}_3$ & $C^{(1,0)}_3+C^{(3,0)}_3+C^{(2,1)}_3 $& $C^{(1,0)}_3+C^{(3,0)}_3+C^{(1,2)}_3$ \\
       \hline
       $C^{(3,0)}_3$& $C^{(3,0)}_3$ & $3C_1+C^{(1,2)}_3+C^{(2,1)}_3$ & $C^{(1,0)}_3+C^{(1,2)}_3+C^{(2,1)}_3$ & $2C^{(1,0)}_3+C^{(2,0)}_3$ & $C^{(2,0)}_3+C^{(3,0)}_3+C^{(1,2)}_3$ & $C^{(2,0)}_3+C^{(3,0)}_3+C^{(2,1)}_3$ \\
       \hline
       $C^{(1,2)}_3$& $C^{(1,2)}_3$ & $C^{(1,0)}_3+C^{(2,0)}_3+C^{(1,2)}_3$ & $C^{(1,0)}_3+C^{(3,0)}_3+C^{(2,1)}_3$ &  $C^{(2,0)}_3+C^{(3,0)}_3+C^{(1,2)}_3$ &$C^{(2,0)}_3+2C^{(2,1)}_3$ & $3C_1+C^{(1,0)}_3+C^{(3,0)}_3$  \\
       \hline
       $C^{(2,1)}_3$& $C^{(2,1)}_3$ & $C^{(1,0)}_3+C^{(2,0)}_3+C^{(2,1)}_3$ & $C^{(1,0)}_3+C^{(3,0)}_3+C^{(1,2)}_3$ & $C^{(2,0)}_3+C^{(3,0)}_3+C^{(2,1)}_3$ & $3C_1+C^{(1,0)}_3+C^{(3,0)}_3$ & $C^{(2,0)}_3+2C^{(1,2)}_3$ \\
       \hline
    \end{tabular}
    }
  \end{table}
\end{itemize}

$\mathbb{Z}_3$ gauging of $\mathbb{Z}_N \times \mathbb{Z}'_N$ can be realized by higher dimensional field theory and string theory similar to $\mathbb{Z}_2$ gauging of $\mathbb{Z}_N$.
Certain compactifications can lead to $\mathbb{Z}_N \times \mathbb{Z}'_N$.
Under such symmetry, massless modes $\varphi_{k,\ell}$ transform as 
\begin{align}\label{eq:ZN}
    \varphi_{k,\ell} \to a^k \varphi_{k,\ell}
\end{align}
under $\mathbb{Z}_N$, and 
\begin{align}\label{eq:ZN'}
    \varphi_{k,\ell} \to a'^\ell \varphi_{k,\ell}
\end{align}
under $\mathbb{Z}'_N$.
Here, $k$ and $\ell$ may correspond to two independent discrete momenta, winding numbers, and their mixtures in the compact space.

Here, we consider a proper $\mathbb{Z}_3$  orbifolding such that $\mathbb{Z}_3$-invariant mode is written by \footnote{See Refs.~\cite{Abe:2013bca,Abe:2014noa,Kobayashi:2017dyu} for geometrical $\mathbb{Z}_N$ orbifolding of magnetized compactifications.}
\begin{align}
    \phi_{k,\ell} = \varphi_{k,\ell} + \varphi_{-k+\ell,-k} + \varphi_{\ell,k-\ell}.
\end{align}
That corresponds to $C^{(k,\ell)}$.
We study the model with $N=3$ as an illustrating model.
Suppose that the "momentum" $(k,\ell)=(1,0)$ in the compact space corresponds to one of the simple roots of $SU(3)$, $\alpha_1$ and the 
"momentum" $(k,\ell)=(0,1)$ corresponds to the other $\alpha_2$.
Then, the $\mathbb{Z}_3$ orbifolding corresponds to dividing $T^2=R^2/\Lambda_{SU(3)}$ by the $\mathbb{Z}_3$ Coxeter element of $SU(3)$, where $\Lambda_{SU(3)}$ denotes the $SU(3)$ root lattice. 
The fields $\phi_{1,0}$ and $\phi_{2,0}$ corresponding to $C_3^{(1,0)}$ and $C_3^{(2,0)}$ are written as 
\begin{align}
    \phi_{1,0} = \varphi_{1,0} + \varphi_{2,2} + \varphi_{0,1}, \qquad 
    \phi_{2,0} = \varphi_{2,0} + \varphi_{1,1} + \varphi_{0,2},
\end{align}
where the "momenta" $(k,\ell)$ appearing in the right hand sides are 
six non-zero roots of $SU(3)$.
They are $\mathbb{Z}_3$ twist invariant modes \cite{Beye:2014nxa}.

$\mathbb{Z}_3$ gauging of $\mathbb{Z}_N \times \mathbb{Z}'_N$ leads to field-theoretical properties different from $\mathbb{Z}_2$ gauging of $\mathbb{Z}_N$, but similar to $\mathbb{Z}_3$ gauging of $\mathbb{Z}_N$.
Each field $\phi_{k,\ell}$ can correspond to a class $C^{(k,\ell)}$.
They can be complex representations different from $\mathbb{Z}_2$ gauging.
The coupling selection rules can be derived in a way similar to coupling selection rules due to $\mathbb{Z}_2$ gauging to $\mathbb{Z}_N$ symmetries.
That is, if the products of $\prod C^{(k,\ell)}$ includes the identity $e$, the $n$-point coupling $\prod \phi_{k,\ell}$ is allowed.
For example, the two-point coupling $\phi_{k,\ell} \phi_{k,\ell}$ is not always allowed, because $C^{k,\ell}\cdot C^{k,\ell}$ does not include the identity $e$ for generic $(k,\ell)$.
That is different from the coupling selection rules due to $\mathbb{Z}_2$ of $\mathbb{Z}_N$ symmetries.
On the other hand, we find that  $(C^{k,\ell})^3$ always includes the identity.
Thus, the three-point self coupling $(\phi_{k,\ell})^3$ is always allowed.

\subsection{$\Delta(3N^2)$}

In this section, we discuss the relations of $\mathbb{Z}_3$ gauging of the $\mathbb{Z}_N\times \mathbb{Z}'_N$ symmetries to the $\Delta(3N^2)=(\mathbb{Z}_N \times \mathbb{Z}'_N) \rtimes \mathbb{Z}_3$ symmetries.
The classes, $C_1$, $C_1^{s}$, and $C_3^{(k,\ell)}$ in the previous section correspond to the conjugacy classes of $\Delta(3N^2)$, which do not include $b$ \cite{Ishimori:2010au,Kobayashi:2022moq}.\footnote{Here, we follow the notation in Refs.~\cite{Ishimori:2010au,Kobayashi:2022moq}.}
The $\Delta(3N^2)$ group has other conjugacy classes.
When $N/3=$~integer, there exist the following conjugacy classes:
\begin{align}
    B_{N^2/3}^{(1,p,0)}&\equiv \{ba^{p-n-3m}a'^{n}| m=0,1,\cdots,\frac{N-3}{3},n=0,1,\cdots N-1 \},
    \nonumber\\
    B_{N^2/3}^{(2,p,0)}&\equiv \{b^2a^{p-n-3m}a'^{n}| m=0,1,\cdots,\frac{N-3}{3},n=0,1,\cdots N-1 \},
\end{align}
with $p=0,1,2$. 
On the other hand, When $N/3\neq$~integer, there exist the following conjugacy classes:
\begin{align}
    C_{N^2}^{1}&\equiv \{ba^\ell a'^{m}| \ell,m=0,1,\cdots,N-1 \},
    \nonumber\\
    C_{N^2}^{2}&\equiv \{b^2a^\ell a'^{m}| \ell,m=0,1,\cdots N-1 \}.
\end{align}

We discuss the realization of the conjugacy classes $B_{N^2/3}^{(1,p,0)}$ as well as $B_{N^2/3}^{(2,p,0)}$.
We focus on $N=3$.\footnote{The $N=2$ case, i.e. $\Delta(12)\cong A_4$ conjugacy classes, is discussed in Appendix~\ref{app:A4S4}.} 
The other conjugacy classes $C^{k,\ell}$ can be realized by 
the "bulk" modes on $T^2/\mathbb{Z}_3$.
We may need to three twist fields $\sigma_p$ to represent 
$B_{N^2/3}^{(1,p,0)}$.
There are three fixed points on $T^2/\mathbb{Z}_3$, and they are represented by the conjugacy classes of space group $(\theta,p\alpha_1)$ with $p=0,1,2$, where $\theta$ denotes the $\mathbb{Z}_3$ twist.
Thus, the structure of the conjugacy classes $B_{N^2/3}^{(1,p,0)}$ is consistent with the geometrical property of $T^2/\mathbb{Z}_3$ orbifold.
There exist the following $\mathbb{Z}_3$ symmetry:
\begin{align}\label{eq:Delta27-1}
    \sigma_p \to \omega^p \sigma_p,
\end{align}
with $\omega=e^{2\pi i /3}$ and 
the cyclic symmetry:
\begin{align}\label{eq:Delta27-2}
    \sigma_0 \to \sigma_1 \to \sigma_2 \to \sigma_0.
\end{align}
Combining these symmetries, the full symmetry becomes $\Delta(27)$ \cite{Kobayashi:2006wq,Beye:2014nxa}.
The conjugacy class $B_{N^2/3}^{(2,p,0)}$ can be represented by $\sigma^2_p$.

When $N/3=$ integer, introduction of the three twist fields $\sigma_p$ can present the conjugacy classes $B_{N^2/3}^{(1,p,0)}$ as well as $B_{N^2/3}^{(2,p,0)}$. Table \ref{tab:Delta27} shows multiplication rules of whole $\Delta(27)$ conjugacy classes:
\begin{align}
    &C_1 = \{e\},\\
    &C^1_1 = \{aa'^2\},\\
    &C^2_1 = \{a^2a'\},\\
    &C^{(1, 0)}_3 = \{a, a', a^2a'^2\},\\
    &C^{(2, 0)}_3 = \{a^2, a'^2, aa'\},\\
    &B^{(1, 0,0)}_3 = \{b, baa'^2, ba^2a'\},\\
    &B^{(1, 1,0)}_3 = \{ba, ba', ba^2a'^2\},\\
    &B^{(1, 2,0)}_3 = \{ba^2, ba'^2, baa'\},\\
    &B^{(2, 0,0)}_3 = \{b^2, b^2aa'^2, b^2a^2a'\},\\
    &B^{(2, 1,0)}_3 = \{b^2a, b^2a', b^2a^2a'^2\},\\
    &B^{(2, 2,0)}_3 = \{b^2a^2, b^2a'^2, b^2aa'\}.
\end{align}

It was known that the outer automorphism of $\Delta(27)$ is generated by~\cite{Holthausen:2012dk}
\begin{align}
    u_1 &: (aa', b) \rightarrow (aa'b(aa')^{-1}, b^{-1}aa'b)\,,
    \nonumber\\
    u_2 &: (aa', b) \rightarrow (b^{-1}(aa')^{-1}, b^2)\,,
\end{align}
which is isomorphic to $\mathrm{GL}(2,3)\cong Q_8\rtimes S_3$. 
It is defined in terms of 
\begin{align}
&(a_1)^4=(a_3)^2=(a_4)^2=(a_3a_4)^3=e,\quad
(a_2)^2=(a_1)^2,\quad
(a_2)^{-1}a_1a_2=(a_1)^{-1},
\nonumber\\
&a_3a_1a_3=(a_1)^3,\quad
a_3a_2a_3=a_1a_2,\quad
a_4a_1a_4=a_2,\quad
a_4a_2a_4=a_1,
\end{align}
with $a_1=u_1u_2u_1u_2$, $a_2=u_2u_1u_2u_1$, $a_3=u_2u_1(u_2)^2$, and $a_4=u_1$, 
where $\{a_1,a_2\}$ generate $Q_8$ and $\{a_3, a_4\}$ generate $S_3$. Specifically, $u_1$ and $u_2$ act on each of the conjugacy classes such as
\begin{align}
    u_1 : \ &C^1_1 \to C^2_1,\quad C^2_1\to C^1_1,\quad C^{(1,0)}_3\to B^{(2,0,0)}_3,\quad C^{(2,0)}_3\to B^{(1,0,0)}_3,\nonumber\\
    &B^{(1,0,0)}_3\to C^{(2,0)}_3,\quad B^{(1,1,0)}_3\to B^{(2,2,0)}_3,\quad B^{(2,0,0)}_3\to C^{(1,0)}_3,\quad B^{(2,2,0)}_3\to B^{(1,1,0)}_3,
\nonumber\\
    u_2 : \ &C^{(1,0)}_3\to B^{(1,2,0)}_3,\quad C^{(2,0)}_3\to B^{(2,1,0)}_3,\quad B^{(1,0,0)}_3\to B^{(2,0,0)}_3,\quad B^{(1,1,0)}_3\to C^{(2,0)}_3,\nonumber\\
    &B^{(1,2,0)}_3\to B^{(1,1,0)}_3,\quad B^{(2,0,0)}_3\to B^{(1,0,0)}_3,\quad B^{(2,1,0)}_3\to B^{(2,2,0)}_3,\quad B^{(2,2,0)}_3\to C^{(1,0)}_3. 
\end{align}
It turns out that the multiplication rules of $\Delta(27)$ conjugacy classes in Table \ref{tab:Delta27} are invariant under the $\mathrm{GL}(2,3)$ symmetry. 
In addition, there is the $\mathbb{Z}_3 \times \mathbb{Z}'_3$ symmetry. 
One $\mathbb{Z}_3$ symmetry corresponds to Eq.~(\ref{eq:Z3-C}). 
To be consistent, the classes $B_3^{(n,p,0)}$ must also transform as 
\begin{align}
    B_3^{(n,p,0)} \to \omega^{n+p}B_3^{(n,p,0)}.
\end{align}
The other $\mathbb{Z}'_3$ symmetry corresponds to the transformation $b \to \omega b$, i.e,
\begin{align}
    B_3^{(n,p,0)} \to \omega^n B_3^{(n,p,0)}.
\end{align}
Hence, the multiplication rules of $\Delta(27)$ conjugacy classes have the $((\mathbb{Z}_3\times \mathbb{Z}_3) \rtimes Q_8)\rtimes S_3$ symmetry. 

When $N/3\neq$ integer, we may introduce a single twist field to represent the conjugacy class  $C_{N^2}^{1}$, which may correspond to one of $\sigma_p$, or all.
However, its geometrical interpretation is not clear.
At any rate, $\mathbb{Z}_3$ invariant elements of $\Delta(3N^2)$ correspond to the elements in $\mathbb{Z}_3$ gauging of $\mathbb{Z}_N\times \mathbb{Z}'_N$ in the previous section.

 \begin{table}[H]
    \centering
    \caption{Multiplication rules for conjugacy classes of $\Delta(27)$.}
\label{tab:Delta27}
\resizebox{\textwidth}{!}{
    \begin{tabular}{|c||c|c|c|c|c|}
      \hline
       & $C_1$ & $C^1_1$ & $C^2_1$ & $C^{(1, 0)}_3$ & $C^{(2, 0)}_3$ \\
       \hline
       \hline
       $C_1$& $C_1$ & $C^1_1$ & $C^2_1$ & $C^{(1, 0)}_3$ & $C^{(2, 0)}_3$ \\
       \hline
       $C^1_1$& $C^1_1$& $C^2_1$ & $C_1$ & $C^{(1, 0)}_3$ & $C^{(2, 0)}_3$ \\
       \hline
       $C^2_1$& $C^2_1$ &$C_1$  & $C^1_1$ & $C^{(1, 0)}_3$ & $C^{(2, 0)}_3$ \\
       \hline
       $C^{(1, 0)}_3$& $C^{(1, 0)}_3$ & $C^{(1, 0)}_3$ & $C^{(1, 0)}_3$ & $3 C^{(2, 0)}_3$ & $3 C_1 + 3 C^1_1 + 3 C^2_1$ \\
       \hline
       $C^{(2, 0)}_3$& $C^{(2, 0)}_3$ & $C^{(2, 0)}_3$ &$C^{(2, 0)}_3$  & $3 C_1 + 3 C^1_1 + 3 C^2_1$ & $3 C^{(1, 0)}_3$ \\
       \hline
       
       $B^{(1,0,0)}_3$& $B^{(1,0,0)}_3$ & $B^{(1,0,0)}_3$ &$B^{(1,0,0)}_3$  & $3 B^{(1,1,0)}_3$ & $3 B^{(1,2,0)}_3$ \\
       \hline  
       $B^{(1,1,0)}_3$& $B^{(1,1,0)}_3$ & $B^{(1,1,0)}_3$ &$B^{(1,1,0)}_3$  & $3 B^{(1,2,0)}_3$ & $3 B^{(1,0,0)}_3$ \\
       \hline  
       $B^{(1,2,0)}_3$& $B^{(1,2,0)}_3$ & $B^{(1,2,0)}_3$ &$B^{(1,2,0)}_3$  & $3 B^{(1,0,0)}_3$ & $3 B^{(1,1,0)}_3$ \\
       \hline  
       $B^{(2,0,0)}_3$& $B^{(2,0,0)}_3$ & $B^{(2,0,0)}_3$ &$B^{(2,0,0)}_3$  & $3 B^{(2,1,0)}_3$ & $3 B^{(2,2,0)}_3$ \\
       \hline  
       $B^{(2,1,0)}_3$& $B^{(2,1,0)}_3$ & $B^{(2,1,0)}_3$ &$B^{(2,1,0)}_3$  & $3 B^{(2,2,0)}_3$ & $3 B^{(2,0,0)}_3$ \\
       \hline  
       $B^{(2,2,0)}_3$& $B^{(2,2,0)}_3$ & $B^{(2,2,0)}_3$ &$B^{(2,2,0)}_3$  & $3 B^{(2,0,0)}_3$ & $3 B^{(2,1,0)}_3$ \\
       \hline  
       \end{tabular}}
       \resizebox{\textwidth}{!}{
    \begin{tabular}{|c||c|c|c|c|c|c|}
      \hline
      & $B^{(1,0,0)}_3$ & $B^{(1,1,0)}_3$ & $B^{(1,2,0)}_3$& $B^{(2,0,0)}_3$ & $B^{(2,1,0)}_3$ & $B^{(2,2,0)}_3$\\
      \hline
      \hline
      $C_1$&$B^{(1,0,0)}_3$ & $B^{(1,1,0)}_3$ & $B^{(1,2,0)}_3$& $B^{(2,0,0)}_3$ & $B^{(2,1,0)}_3$ & $B^{(2,2,0)}_3$\\
      \hline
      $C^1_1$&$B^{(1,0,0)}_3$&$B^{(1,1,0)}_3$&$B^{(1,2,0)}_3$&$B^{(2,0,0)}_3$&$B^{(2,1,0)}_3$&$B^{(2,2,0)}_3$\\
      \hline
      $C^2_1$&$B^{(1,0,0)}_3$&$B^{(1,1,0)}_3$&$B^{(1,2,0)}_3$&$B^{(2,0,0)}_3$&$B^{(2,1,0)}_3$&$B^{(2,2,0)}_3$\\
      \hline
      $C^{(1,0)}_3$&$3B^{(1,1,0)}_3$&$3B^{(1,2,0)}_3$&$3B^{(1,0,0)}_3$&$3B^{(2,1,0)}_3$&$3B^{(2,2,0)}_3$&$3B^{(2,0,0)}_3$\\
      \hline
      $C^{(2,0)}_3$&$3B^{(1,2,0)}_3$&$3B^{(1,0,0)}_3$&$3B^{(1,1,0)}_3$&$3B^{(2,2,0)}_3$&$3B^{(2,0,0)}_3$&$3B^{(2,1,0)}_3$\\
      \hline
      $B^{(1,0,0)}_3$ & $3 B^{(2,0,0)}_3$ & $3 B^{(2,1,0)}_3$ & $3 B^{(2,2,0)}_3$&$3 C_1 + 3 C^1_1 + 3 C^2_1$&$3C^{(1,0)}_3$&$3C^{(2,0)}_3$\\
      \hline
      $B^{(1,1,0)}_3$ & $3 B^{(2,1,0)}_3$ & $3 B^{(2,2,0)}_3$ & $3 B^{(2,0,0)}_3$&$3C^{(1,0)}_3$&$3C^{(2,0)}_3$&$3 C_1 + 3 C^1_1 + 3 C^2_1$\\
      \hline
      $B^{(1,2,0)}_3$ & $3 B^{(2,2,0)}_3$ & $3 B^{(2,0,0)}_3$ & $3 B^{(2,1,0)}_3$&$3C^{(2,0)}_3$&$3 C_1 + 3 C^1_1 + 3 C^2_1$&$3C^{(1,0)}_3$\\
      \hline
      $B^{(2,0,0)}_3$ & $3 C_1 + 3 C^1_1 + 3 C^2_1$ & $3 C^{(1, 0)}_3$ & $3 C^{(2, 0)}_3$&$3 B^{(1,0,0)}_3$ & $3 B^{(1,1,0)}_3$ & $3 B^{(1,2,0)}_3$\\
      \hline
      $B^{(2,1,0)}_3$ & $3 C^{(1, 0)}_3$ & $3 C^{(2, 0)}_3$ & $3 C_1 + 3 C^1_1 + 3 C^2_1$&$3 B^{(1,1,0)}_3$ & $3 B^{(1,2,0)}_3$ & $3 B^{(1,0,0)}_3$\\
      \hline
      $B^{(2,2,0)}_3$ & $3 C^{(2, 0)}_3$ & $3 C_1 + 3 C^1_1 + 3 C^2_1$ & $3 C^{(1, 0)}_3$&$3 B^{(1,2,0)}_3$ & $3 B^{(1,0,0)}_3$ & $3 B^{(1,1,0)}_3$\\
      \hline
    \end{tabular}}
  \end{table}

\section{$\mathbb{Z}_3$ gauging of $\mathbb{Z}_N\times \mathbb{Z}'_N \times \mathbb{Z}''_N$ and $\Sigma(3N^3)$}
\label{sec:Z3NNN}

We can extend $\mathbb{Z}_3$ gauging of $\mathbb{Z}_N \times \mathbb{Z}'_N$ to $\mathbb{Z}_3$ gauging of $\mathbb{Z}_N \times \mathbb{Z}'_N \times \mathbb{Z}''_N$.
The generators of $\mathbb{Z}_N\times \mathbb{Z}'_N\times \mathbb{Z}''_N$ are chosen as
\begin{align}
    &a^N = a'^N = a''^N = e,\notag\\
    &aa' = a'a,\quad a'a'' = a''a',\quad a''a = aa''.
\end{align}
Let us consider the outer automorphism of $\mathbb{Z}_N\times \mathbb{Z}'_N\times \mathbb{Z}''_N$ generated by $b$:
\begin{align}
    &b^3 = e,\quad 
    b^2ab = a'',\quad b^2a'b = a,\quad b^2a''b = a',\notag\\
    &b (a^m a'^n a''^\ell) b^{-1} = a^\ell a'^m a''^n,\quad 
    b^2 (a^m a'^n a''^\ell) b^{-2} = a^n a'^\ell a''^m.
\end{align}
Then, one can define the $\mathbb{Z}_3$-invariant classes:  
\begin{align}
    C^{(k,\ell,m)}=\{ b^n a^k a'^\ell a''^m b^{-n} ~|~n=0,1,2 \}.
\end{align}
They are written more explicitly as 
\begin{align}
    &C_1 = \{ e \},\\
    &C^{s}_1 = \{ (aa'a'')^s \},\\
    &C^{(k,\ell,m)}_3 = \{ a^ka'^\ell a''^m,a^\ell a'^ma''^k,a^ma'^ka''^\ell \},
  \end{align}
with $(k,\ell,m) \neq (0,0,0),(s,s,s)$. These classes satisfy the following fusion rules: 
  \begin{align}
    &C_1\cdot C_1 = C_1,   \\
    &C_1\cdot C^s_1 = C^s_1,   \\
    &C_1\cdot C^{(k,\ell,m)}_3 = C^{(k,\ell,m)}_3,   \\
    &C^s_1\cdot C^t_1 = C^{s+t}_1,   \\
    &C^s_1\cdot C^{(k,\ell,m)}_3 = C^{(k+s,\ell+s,m+s)}_3,  \\
    &C^{(k,\ell,m)}_3\cdot C^{(p,q,r)}_3 = C^{(k+p,\ell+q,m+r)}_3 + C^{(k+r,\ell+p,m+q)}_3 + C^{(k+q,\ell+r,m+p)}_3.  
  \end{align}
    Note that if $C^{0}_1$ and $C^{(s,s,s)}_3$ appear in the right hand side in the above equations, we have to replace them by
  \begin{align}
    &C^0_1 \Rightarrow C_1,\quad C^{(s,s,s)}_3 \Rightarrow  3C^s_1. 
  \end{align}
  For concreteness, we present the multiplication rules of the classes for the $N=2$ case in Table~\ref{tab:Z3gaugingZ2Z2Z2}, from which 
  there exists the $\mathbb{Z}_2$ symmetry associated with $C_1^1 \rightarrow - C_1^1$ and $C^{(1,0,0)}_3\rightarrow - C^{(1,0,0)}_3$.
  \begin{table}[H]
    \centering
    \caption{Multiplication rules of classes for $\mathbb{Z}_3$ gauging of $\mathbb{Z}_2\times \mathbb{Z}'_2\times \mathbb{Z}''_2$.}
    \label{tab:Z3gaugingZ2Z2Z2}
    \begin{tabular}{|c||c|c|c|c|}
    \hline
       &$C_1$ &$C^1_1$ &$C^{(1,0,0)}_3$ &$C^{(1,1,0)}_3$  \\
       \hline\hline
       $C_1$&$C_1$ &$C^1_1$ & $C^{(1,0,0)}_3$&$C^{(1,1,0)}_3$ \\
       \hline
       $C^1_1$&$C^1_1$ &$C_1$ &$C^{(1,1,0)}_3$ &$C^{(1,0,0)}_3$ \\
       \hline
       $C^{(1,0,0)}_3$&$C^{(1,0,0)}_3$ &$C^{(1,1,0)}_3$ &$3C_1+2C^{(1,1,0)}_3$ &$3C^1_1+2C^{(1,0,0)}_3$ \\
       \hline
       $C^{(1,1,0)}_3$&$C^{(1,1,0)}_3$ &$C^{(1,0,0)}_3$ &$3C^1_1+2C^{(1,0,0)}_3$  &$3C_1+2C^{(1,1,0)}_3$ \\
       \hline
    \end{tabular}
  \end{table}

$\mathbb{Z}_3$ gauging of $\mathbb{Z}_N \times \mathbb{Z'}_N \times \mathbb{Z}''_N$ can be realized by higher dimensional field theory as well as string theory similar to $\mathbb{Z}_3$ gauging of $\mathbb{Z}_N \times \mathbb{Z}'_N$.
First, we consider the compactification leading to $\mathbb{Z}_N \times \mathbb{Z}'_N \times \mathbb{Z}''_N$.
Under such symmetry, massless modes $\varphi_{k,\ell,m}$ transform as
\begin{align}
    \varphi_{k,\ell,m} \to a^k a'^\ell a''^m \varphi_{k,\ell,m},
\end{align}
where $k, \ell, m$ may correspond to three independent discrete momenta, winding numbers, and their mixtures in the compact space.
Then, we divide this compact space further by $A_3 \simeq \mathbb{Z}_3$ permutation. 
The $\mathbb{Z}_3$-invariant mode is written by 
\begin{align}\label{eq:field-Z3-ZN-3}
    \phi_{k,\ell,m}=\varphi_{k,\ell,m}+\varphi_{\ell, m, k} + \varphi_{m, k, \ell}.
\end{align}
That corresponds to $C^{(k,\ell,m)}$.

The above classes $C^{(k,\ell,m)}$ correspond to the conjugacy classes without $b$ of $(\mathbb{Z}_N \times \mathbb{Z}'_N \times \mathbb{Z}''_N) \rtimes \mathbb{Z}_3=\Sigma (3M^3)$.
That is, $C^{(k,\ell,m)}$ is $\mathbb{Z}_3$ invariant ones. 
Table \ref{tab:Sigma(24)} shows multiplication rules of whole $\Sigma(24)$ conjugacy classes:
\begin{align*}
    &C_1 = \{e\},\\
    &C^1_1 = \{aa'a''\},\\
    &C^{(1,1,0)}_3 = \{aa', a'a'', a''a\},\\
    &C^{(1,0,0)}_3 = \{a, a', a''\},\\
    &B^{(1,0,0,0)}_4 = \{b, baa', ba'a'', ba''a\},\\
    &B^{(1, 1,0,0)}_4 = \{ba, ba', ba'', baa'a''\},\\
    &B^{(2, 0,0,0)}_4 = \{b^2, b^2aa', b^2a'a'', b^2a''a\},\\
    &B^{(2, 1,0,0)}_4 = \{b^2a, b^2a', b^2a'', b^2aa'a''\}.
\end{align*}
We find that there exists the $\mathbb{Z}_2$ symmetry:
\begin{align}
    C_1^1 \rightarrow - C_1^1,\quad
    C^{(1,0,0)}_3\rightarrow - C^{(1,0,0)}_3,\quad
    B^{(1,1,0,0)}_4\rightarrow - B^{(1,1,0,0)}_4,\quad
    B^{(2,1,0,0)}_4\rightarrow - B^{(2,1,0,0)}_4,
\end{align}
and the others are unchanged.

\begin{table}[H]
  \centering
  \caption{Multiplication rules for conjugacy classes of $\Sigma(24)$.}
  \label{tab:Sigma(24)}
  \resizebox{\textwidth}{!}{
  \begin{tabular}{|c||c|c|c|c|c|c|c|c|}
    \hline
    & $C_1$ & $C^1_1$ & $C^{(1,1,0)}_3$ & $C^{(1,0,0)}_3$ & $B^{(1,0,0,0)}_4$ & $B^{(1,1,0,0)}_4$ & $B^{(2,0,0,0)}_4$ & $B^{(2,1,0,0)}_4$\\
    \hline
    \hline
    $C_1$ & $C_1$ & $C^1_1$ & $C^{(1,1,0)}_3$ & $C^{(1,0,0)}_3$ & $B^{(1,0,0,0)}_4$ & $B^{(1,1,0,0)}_4$ & $B^{(2,0,0,0)}_4$ & $B^{(2,1,0,0)}_4$\\
    \hline
    $C^1_1$ & $C^1_1$ & $C_1$ & $C^{(1,0,0)}_3$ & $C^{(1,1,0)}_3$ & $B^{(1,1,0,0)}_4$ & $B^{(1,0,0,0)}_4$ & $B^{(2,1,0,0)}_4$ & $B^{(2,0,0,0)}_4$\\
    \hline
    $C^{(1,1,0)}_3$ & $C^{(1,1,0)}_3$ & $C^{(1,0,0)}_3$ & $3 C_1 + 2 C^{(1,1,0)}_3$ & $3 C^1_1 + 2 C^{(1,0,0)}_3$ & $3 B^{(1,0,0,0)}_4$ & $3 B^{(1,1,0,0)}_4$ & $3 B^{(2,0,0,0)}_4$ & $3 B^{(2,1,0,0)}_4$\\
    \hline
    $C^{(1,0,0)}_3$ & $C^{(1,0,0)}_3$ & $C^{(1,1,0)}_3$ & $3 C^1_1 + 2 C^{(1,0,0)}_3$ & $3 C_1 + 2 C^{(1,1,0)}_3$ & $3 B^{(1,1,0,0)}_4$ & $3 B^{(1,0,0,0)}_4$ & $3 B^{(2,1,0,0)}_4$ & $3 B^{(2,0,0,0)}_4$\\
    \hline
    $B^{(1,0,0,0)}_4$ & $B^{(1,0,0,0)}_4$ & $B^{(1,1,0,0)}_4$ & $3 B^{(1,0,0,0)}_4$ & $3 B^{(1,1,0,0)}_4$ & $4 B^{(2,0,0,0)}_4$ & $4 B^{(2,1,0,0)}_4$ & $4 C_1 + 4 C^{(1,1,0)}_3$ & $4 C^1_1 + 4 C^{(1,0,0)}_3$\\
    \hline
    $B^{(1,1,0,0)}_4$ & $B^{(1,1,0,0)}_4$ & $B^{(1,0,0,0)}_4$ & $3 B^{(1,1,0,0)}_4$ & $3 B^{(1,0,0,0)}_4$ & $4 B^{(2,1,0,0)}_4$ & $4 B^{(2,0,0,0)}_4$ & $4 C^1_1 + 4 C^{(1,0,0)}_3$ & $4 C_1 + 4 C^{(1,1,0)}_3$\\
    \hline
    $B^{(2,0,0,0)}_4$ & $B^{(2,0,0,0)}_4$ & $B^{(2,1,0,0)}_4$ & $3 B^{(2,0,0,0)}_4$ & $3 B^{(2,1,0,0)}_4$ & $4 C_1 + 4 C^{(1,1,0)}_3$ & $4 C^1_1 + 4 C^{(1,0,0)}_3$ & $4 B^{(1,0,0,0)}_4$ & $4 B^{(1,1,0,0)}_4$\\
    \hline
    $B^{(2,1,0,0)}_4$ & $B^{(2,1,0,0)}_4$ & $B^{(2,0,0,0)}_4$ & $3 B^{(2,1,0,0)}_4$ & $3 B^{(2,0,0,0)}_4$ & $4 C^1_1 + 4 C^{(1,0,0)}_3$ & $4 C_1 + 4 C^{(1,1,0)}_3$ & $4 B^{(1,1,0,0)}_4$ & $4 B^{(1,0,0,0)}_4$\\
    \hline
  \end{tabular}
  }
\end{table}

The conjugacy classes without $b$ can be realized by the "bulk" fields as Eq.~(\ref{eq:field-Z3-ZN-3}).
If the above $A_3 \simeq \mathbb{Z}_3$ permutation would induce twist fields, the conjugacy classes with $b$ may be realized by such twist fields.

\section{$S_3$ gauging of $\mathbb{Z}_N\times \mathbb{Z}'_N$ and $\Delta(6N^2)$}
\label{sec:S3gauge}

\subsection{$S_3$ gauging of $\mathbb{Z}_N\times \mathbb{Z}'_N$}

In this section, we analyze another gauging of $\mathbb{Z}_N\times \mathbb{Z}'_N$, i.e., 
$S_3$ gauging. 
The generators of $\mathbb{Z}_N \times \mathbb{Z}'_N$   are the same as in Sec.~\ref{subsec:Z3gauge}, but we consider the outer automorphisms of $\mathbb{Z}_N\times \mathbb{Z}'_N$ generated by $b$ and $c$:
\begin{align}
    &b^3 = c^2 = (bc)^2 = e,\notag\\
    &bab^{-1} = a^{-1}a'^{-1},\quad ba'b^{-1} = a,\notag\\
    &cac^{-1} = a'^{-1},\quad ca'c^{-1} = a^{-1},\notag\\
    &b (a^\ell a'^m) b^{-1} = a^{-\ell + m} a'^{-\ell},\notag \\
    &b^2 (a^\ell a'^m) b^{-2} = a^{-m} a'^{\ell - m},\notag \\
    &c (a^\ell a'^m) c^{-1} = a^{-m} a'^{-\ell},\notag \\
    &b c (a^\ell a'^m) (b c)^{-1} = a^{m - \ell} a'^{m},\notag \\
    &b^2 c (a^\ell a'^m) (b^2 c)^{-1} = a^{-\ell + m} a'^{-\ell}.
\end{align}
Then, one can define the $S_3$-invariant class:
  \begin{align}
\label{eq:S3classofZNZN}
C^{(k,\ell)} &\equiv \{b^nc^m a^k a'^\ell (b^nc^m)^{-1} ~|~ n=0,1,2, ~m=0,1 \} \notag \\
& = \{ a^ka'^\ell, a^{-k+\ell}a'^{-k}, a^{-\ell}a'^{k-\ell},a^{-\ell}a'^{-k}, a^{-k+\ell}a'^\ell, a^ka'^{k-\ell} \}.
  \end{align}
In this context of $S_3$ gauging of the  $\mathbb{Z}_N \times \mathbb{Z}'_N$ symmetry, the classes obey the following fusion rule:
  \begin{align}
&C^{(k,\ell)}\cdot C^{(m,n)} = C^{(k+m,\ell+n)}+C^{(k-m+n,\ell-m)}+C^{(k-n,\ell+m-n)}\notag\\
    &\qquad\qquad\qquad+C^{(k-n,\ell-m)}+C^{(k-m+n,\ell+n)}+ C^{(k+m,\ell+m-n)}.
  \end{align}
In general, the class $C^{(k,\ell)}$ includes six elements, as shown in Eq.~\eqref{eq:S3classofZNZN}, but obviously for $k=\ell=0$, the class $C^{(0,0)}$ includes only one element $a^0a'^0=e$. 
In the following, we consider two cases depending on the value of $N$:

\begin{itemize}
  \item $N/3\neq \ $integer 

When $N/3$ is not an integer, we have three classes:  
  \begin{align}
    &C_1 = \{ e \},\\
    &C^{s}_3 = \{ a^sa'^{-s}, a^{-2s}a'^{-s}, a^sa'^{2s} \},\\
    &C^{(k,\ell)}_6 = \{ a^ka'^\ell, a^{-k+\ell}a'^{-k}, a^{-\ell}a'^{k-\ell},a^{-\ell}a'^{-k}, a^{-k+\ell}a'^\ell, a^ka'^{k-\ell} \},
  \end{align}
  with $l\neq -k$, and their fusion rules are given by
  \begin{align}
    &C_1\cdot C_1 = C_1,\\
    &C_1\cdot C^s_3 = C^s_3,\\
    &C_1\cdot C^{(k,\ell)}_6 = C^{(k,\ell)}_6,\\
    &C^s_3\cdot C^t_3 = C^{s+t}_3+C^{(s-2t,-s-t)}_6,\\
    &C^s_3\cdot C^{(k,\ell)}_6 = C^{(s+k,-s+\ell)}_6+C^{(s-k+\ell,-s-k)}_6+C^{(s-\ell,-s+k-\ell)}_6, \\
    &C^{(k,\ell)}_6\cdot  C^{(m,n)}_6 = C^{(k+m,\ell+n)}_6+C^{(k-m+n,\ell-m)}_6+C^{(k-n,\ell+m-n)}_6\notag\\
    &\qquad\qquad\qquad+C^{(k-n,\ell-m)}_6+C^{(k-m+n,\ell+n)}_6+ C^{(k+m,\ell+m-n)}_6,
    \end{align}
  Note that if $C^{0}_3$, $C^{(0,0)}_6$ and $C^{(k,-k)}_6$ appear in the right hand side in the above equations, we have to replace them by
  \begin{align}
    &C^{0}_3 \Rightarrow 3C_1,\quad
    C^{(0,0)}_6 \Rightarrow 6C_1,\quad
    C^{(k,-k)}_6 \Rightarrow 2C^{k}_3.
  \end{align}
  
  \item $N/3= \ $integer

  When $N/3$ is an integer, we have four classes:    
  \begin{align}
  \begin{array}{ll}
    C_1 = \{ e \},&\\
    C^{s}_1 = \{ a^sa'^{-s} \}& (s = N/3,2N/3),\\
    C^{s}_3 = \{ a^sa'^{-s}, a^{-2s}a'^{-s}, a^sa'^{2s} \}&(s\neq N/3, 2N/3),\\
    C^{(k,\ell)}_6 = \{ a^ka'^\ell, a^{-k+\ell}a'^{-k}, a^{-\ell}a'^{k-\ell},a^{-\ell}a'^{-k}, a^{-k+\ell}a'^\ell, a^ka'^{k-\ell} \}& (\ell\neq -k).
  \end{array}
  \end{align}
  Their fusion rules are given by 
    \begin{align}
    &C_1\cdot C_1 = C_1,\\
    &C_1\cdot C^{s}_1 = C^{s}_1,\\
    &C_1\cdot C^{s}_3 = C^{s}_3,\\
    &C_1\cdot C^{(k,\ell)}_6 = C^{(k,\ell)}_6,\\
    &C^s_1\cdot C^{s'}_1 = 
  \left\{ \,
    \begin{aligned}
    & C^{2s}_1\quad (s=s') \\
    & C_1\quad (s\neq s') \\
    \end{aligned}
\right.,\\
    &C^s_1\cdot C^t_3 = C^{(s+t)}_3,\\
    &C^s_1\cdot C^{(k,\ell)}_6 = C^{(s+k,-s+\ell)}_6,\\
    &C^s_3\cdot C^t_3 = C^{s+t}_3+C^{(s-2t,-s-t)}_6,\\
    &C^s_3\cdot C^{(k,\ell)}_6 = C^{(s+k,-s+\ell)}_6+C^{(s-k+\ell,-s-k)}_6+C^{(s-\ell,-s+k-\ell)}_6 ,\\
    &C^{(k,\ell)}_6\cdot C^{(m,n)}_6 = C^{(k+m,\ell+n)}_6+C^{(k-m+n,\ell-m)}_6+C^{(k-n,\ell+m-n)}_6\notag\\
    &\qquad\qquad\qquad+C^{(k-n,\ell-m)}_6+C^{(k-m+n,\ell+n)}_6+ C^{(k+m,\ell+m-n)}_6.
  \end{align}
    Note that if $C^{0}_3$, $C^{(0,0)}_6$, $C^{(N/3,2N/3)}_6$, $C^{(2N/3,N/3)}_6$ and $C^{(k,-k)}_6$ appear in the right hand side in the above equations, we have to replace them by
  \begin{align}
    &C^{0}_3 \Rightarrow 3C_1,\quad
    C^{(0,0)}_6 \Rightarrow 6C_1,\quad
    C^{(N/3,2N/3)}_6 \Rightarrow 6C^{N/3}_1,\quad
    C^{(2N/3,N/3)}_6 \Rightarrow 6C^{2N/3}_1,\notag\\
    &C^{(k,-k)}_6 \Rightarrow 2C^{k}_3.
  \end{align}
  \end{itemize}
For concreteness, we present the multiplication rules of the classes for cases with $N=2$ in Table~\ref{tab:S3gaugingZ2Z2} and $N=3$ in Table~\ref{tab:S3gaugingZ3Z3}.\footnote{Multiplication rules of $S_3$ gauging of $\mathbb{Z}_3 \times \mathbb{Z}'_3$ are similar to the $\mathbb{Z}_3$ Tambara-Yamagami fusion rules. See Appendix \ref{app:futher}.} It turns out that 
there exists the $S_2$ symmetry associated with $C_1^1 \leftrightarrow C_1^2$ for $N=3$. 
\begin{table}[H]
  \centering
  \caption{Multiplication rules of classes for $S_3$ gauging of $\mathbb{Z}_2\times\mathbb{Z}'_2$.}
  \label{tab:S3gaugingZ2Z2}
  \begin{tabular}{|c||c|c|}
  \hline
     & $C_1$&$C^1_3$ \\
     \hline\hline
     $C_1$& $C_1$ & $C^1_3$ \\
     \hline
     $C^1_3$& $C^1_3$ & $3C_1+2C^1_3$ \\ 
     \hline
  \end{tabular}
\end{table}
\begin{table}[H]
  \centering
  \caption{Multiplication rules of classes for $S_3$ gauging of $\mathbb{Z}_3\times\mathbb{Z}'_3$.}
  \label{tab:S3gaugingZ3Z3}
  \begin{tabular}{|c||c|c|c|c|}
  \hline
     &$C_1$ &$C^1_1$ &$C^2_1$ &$C^{(1,0)}_6$ \\
     \hline
     \hline
     $C_1$&$C_1$ &$C^1_1$ &$C^2_1$ &$C^{(1,0)}_6$ \\
     \hline
     $C^1_1$& $C^1_1$&$C^2_1$ & $C_1$& $C^{(1,0)}_6$\\
     \hline
     $C^2_1$& $C^2_1$&$C_1$ &$C^1_1$ & $C^{(1,0)}_6$\\
     \hline
     $C^{(1,0)}_6$&$C^{(1,0)}_6$ & $C^{(1,0)}_6$&$C^{(1,0)}_6$ &$6C_1+6C^1_1+6C^2_1+3C^{(1,0)}_6$ \\
     \hline
  \end{tabular}
\end{table}

$S_3$ gauging of $(\mathbb{Z}_N \times \mathbb{Z}'_N )$ can be realized by higher dimensional theory in a way similar to $\mathbb{Z}_3$ gauging of $(\mathbb{Z}_N \times \mathbb{Z}'_N)$.
The $(\mathbb{Z}_N \times \mathbb{Z}'_N)$ transformation is the same as Eqs.(\ref{eq:ZN}) and (\ref{eq:ZN'}).
Then, we consider a proper $S_3$ orbifolding.
For example, when $N=3$ and $T^2$ is the $T^2=R^2/\Lambda_{SU(2)}$ in the previous section, we use a combination of the $\mathbb{Z}_3$ Coxeter element and $\mathbb{Z}_2$ reflection.

\subsection{$\Delta(6N^2)$}

In this section, we discuss the relations of $S_3$ gauging of the $\mathbb{Z}_N\times \mathbb{Z}'_N$ symmetries to the $\Delta(6N^2)=(\mathbb{Z}_N \times \mathbb{Z}'_N) \rtimes S_3$ symmetries.
The classes, $C_1$, $C_1^s$, $C_3^{s}$, and $C_6^{(k,\ell)}$ in the previous section correspond to the conjugacy classes of $\Delta(6N^2)$, which do not include $b$ and $c$ \cite{Ishimori:2010au,Kobayashi:2022moq}.\footnote{Here, we follow the notation in Refs.~\cite{Ishimori:2010au,Kobayashi:2022moq}.}

The full conjugacy classes of $\Delta(6N^2)$ include additional classes. 
For a concrete example, let us consider $\Delta(54)$ group whose the conjugacy classes are listed as\footnote{The multiplication rules of $\Delta(24)\cong S_4$ conjugacy classes are discussed in Appendix~\ref{app:A4S4}.}
\begin{align}
&C_1=\{ e \}\,,\\
&C^1_1=\{ aa'^2 \}\,,\\
&C^2_1=\{ a^2a' \}\,,\\
&C^{(1,0)}_6=\{ a,a',a^2a'^2,a^2,a'^2,aa' \}\,,\\
&B^{(0,0)}_6 = \{ b,ba^2a',baa'^2,b^2,b^2a^2a',b^2aa'^2 \}\,,\\
&B^{(1,0)}_6 = \{ ba,ba',ba^2a'^2,b^2a^2,b^2a'^2,b^2aa' \}\,,\\
&B^{(2,0)}_6 = \{ ba^2,baa',ba'^2,b^2a,b^2a',b^2a^2a'^2 \}\,,\\
&B^{(m)}_9 = \{ ca^{m+p}a'^p, b^2ca^{-m}a'^{-m-p},bca^{-p}a'^m | p=0,1,2\},\quad  (m= 0,1,2)\,.
\end{align}
Note that there exists the outer automorphism generated by~\cite{Fallbacher:2015rea}
\begin{align}
u_1&: (a a', b, c) \rightarrow (aa',aa'baa', c)\,,
\nonumber\\
u_2&: (aa', b, c) \rightarrow (aa'b^2aa', b, c)\,,
\end{align}
which is isomorphic to $S_4$, i.e., $(u_1)^3 = (u_2)^2 = (u_1^{-1}\circ u_2)^4 = e$. 
Here, $u_1$ and $u_2$ act on each of the conjugacy classes such as
\begin{align}
    u_1 : 
    \ &B^{(0,0)}_6 \to B^{(1,0)}_6,
    \quad 
    B^{(1,0)}_6 \to B^{(2,0)}_6,
    \quad
    B^{(2,0)}_6 \to B^{(0,0)}_6,
\nonumber\\
    u_2 : 
    \ &C^1_1 \to C^2_1,\quad
    C^2_1 \to C^1_1,\quad
    C^{(1,0)}_6 \to B^{(2,0)}_6,\quad
    B^{(2,0)}_6 \to C^{(1,0)}_6,\quad
    B^{(1)}_9 \to B^{(2)}_9,\nonumber\\
    &B^{(2)}_9 \to B^{(1)}_9.
\end{align}
It turns out that the multiplication rules of $\Delta(54)$ conjugacy classes in Table \ref{tab:Delta54} are invariant under the $S_4$ symmetry.

\begin{table}[H]
  \centering
  \caption{Multiplication rules for conjugacy classes of $\Delta(54)$.}
  \label{tab:Delta54}
  \resizebox{\textwidth}{!}{
  \begin{tabular}{|c||c|c|c|c|c|c|c|}
  \hline
     &$C_1$ &$C^1_1$ &$C^2_1$ &$C^{(1,0)}_6$&$B^{(0,0)}_6$&$B^{(1,0)}_6$&$B^{(2,0)}_6$ \\
     \hline
     \hline
     $C_1$&$C_1$ &$C^1_1$ &$C^2_1$ &$C^{(1,0)}_6$&$B^{(0,0)}_6$&$B^{(1,0)}_6$&$B^{(2,0)}_6$ \\
     \hline
     $C^1_1$& $C^1_1$&$C^2_1$ & $C_1$& $C^{(1,0)}_6$&$B^{(0,0)}_6$&$B^{(1,0)}_6$&$B^{(2,0)}_6$\\
     \hline
     $C^2_1$& $C^2_1$&$C_1$ &$C^1_1$ & $C^{(1,0)}_6$&$B^{(0,0)}_6$&$B^{(1,0)}_6$&$B^{(2,0)}_6$\\
     \hline
     $C^{(1,0)}_6$&$C^{(1,0)}_6$ & $C^{(1,0)}_6$&$C^{(1,0)}_6$ &$6C_1+6C^1_1+6C^2_1+3C^{(1,0)}_6$&$3B^{(1,0)}_6+3B^{(2,0)}_6$&$3B^{(0,0)}_6+3B^{(2,0)}_6$&$3B^{(0,0)}_6+3B^{(1,0)}_6$\\
     \hline
     $B^{(0,0)}_6$&$B^{(0,0)}_6$&$B^{(0,0)}_6$&$B^{(0,0)}_6$&$3B^{(1,0)}_6+3B^{(2,0)}_6$&$3B^{(0,0)}_6+6C_1+6C^1_1+6C^2_1$&$3B^{(2,0)}_{6}+3C^{(1,0)}_6$&$3B^{(1,0)}_{6}+3C^{(1,0)}_6$\\
     \hline
     $B^{(1,0)}_6$&$B^{(1,0)}_6$&$B^{(1,0)}_6$&$B^{(1,0)}_6$&$3B^{(0,0)}_6+3B^{(2,0)}_6$&$3B^{(2,0)}_6+3C^{(1,0)}_6$&$3B^{(1,0)}_6+6C_1+6C^1_1+6C^2_1$&$3B^{(0,0)}_6+3C^{(1,0)}_6$\\
     \hline
     $B^{(2,0)}_6$&$B^{(2,0)}_6$&$B^{(2,0)}_6$&$B^{(2,0)}_6$&$3B^{(0,0)}_6+3B^{(1,0)}_6$&$3B^{(1,0)}_{6}+3C^{(1,0)}_6$&$3B^{(0,0)}_6+3C^{(1,0)}_6$&$3B^{(2,0)}_6+6C_1+6C^1_1+6C^2_1$\\
     \hline
     $B^{(0)}_9$&$B^{(0)}_9$&$B^{(2)}_9$&$B^{(1)}_9$&$2B^{(0)}_9+2B^{(1)}_9+2B^{(2)}_9$&$2B^{(0)}_9+2B^{(1)}_9+2B^{(2)}_9$&$2B^{(0)}_9+2B^{(1)}_9+2B^{(2)}_9$&$2B^{(0)}_9+2B^{(1)}_9+2B^{(2)}_9$\\
     \hline
     $B^{(1)}_9$&$B^{(1)}_9$&$B^{(0)}_9$&$B^{(2)}_9$&$2B^{(0)}_9+2B^{(1)}_9+2B^{(2)}_9$&$2B^{(0)}_9+2B^{(1)}_9+2B^{(2)}_9$&$2B^{(0)}_9+2B^{(1)}_9+2B^{(2)}_9$&$2B^{(0)}_9+2B^{(1)}_9+2B^{(2)}_9$\\
     \hline
     $B^{(2)}_9$&$B^{(2)}_9$&$B^{(1)}_9$&$B^{(0)}_9$&$2B^{(0)}_9+2B^{(1)}_9+2B^{(2)}_9$&$2B^{(0)}_9+2B^{(1)}_9+2B^{(2)}_9$&$2B^{(0)}_9+2B^{(1)}_9+2B^{(2)}_9$&$2B^{(0)}_9+2B^{(1)}_9+2B^{(2)}_9$\\
     \hline
  \end{tabular}
  }
    \resizebox{\textwidth}{!}{
  \begin{tabular}{|c||c|c|c|}
  \hline
     &$B^{(0)}_9$&$B^{(1)}_9$&$B^{(2)}_9$ \\
     \hline
     \hline
     $C_1$&$B^{(0)}_9$&$B^{(1)}_9$&$B^{(2)}_9$ \\
     \hline
     $C^1_1$&$B^{(2)}_9$&$B^{(0)}_9$&$B^{(1)}_9$\\
     \hline
     $C^2_1$&$B^{(1)}_9$&$B^{(2)}_9$&$B^{(0)}_9$\\
     \hline
     $C^{(1,0)}_6$&$2B^{(0)}_9+2B^{(1)}_9+2B^{(2)}_9$&$2B^{(0)}_9+2B^{(1)}_9+2B^{(2)}_9$&$2B^{(0)}_9+2B^{(1)}_9+2B^{(2)}_9$ \\
     \hline
     $B^{(0,0)}_6$&$2B^{(0)}_9+2B^{(1)}_9+2B^{(2)}_9$&$2B^{(0)}_9+2B^{(1)}_9+2B^{(2)}_9$&$2B^{(0)}_9+2B^{(1)}_9+2B^{(2)}_9$\\
     \hline
     $B^{(1,0)}_6$&$2B^{(0)}_9+2B^{(1)}_9+2B^{(2)}_9$&$2B^{(0)}_9+2B^{(1)}_9+2B^{(2)}_9$&$2B^{(0)}_9+2B^{(1)}_9+2B^{(2)}_9$\\
     \hline
     $B^{(2,0)}_6$&$2B^{(0)}_9+2B^{(1)}_9+2B^{(2)}_9$&$2B^{(0)}_9+2B^{(1)}_9+2B^{(2)}_9$&$2B^{(0)}_9+2B^{(1)}_9+2B^{(2)}_9$\\
     \hline
     $B^{(0)}_9$&$9C_1+3C^{(1,0)}_6+3B^{(0,0)}_6+3B^{(1,0)}_6+3B^{(2,0)}_6$&$9C^2_1+3C^{(1,0)}_6+3B^{(0,0)}_6+3B^{(1,0)}_6+3B^{(2,0)}_6$&$9C^1_1+3C^{(1,0)}_6+3B^{(0,0)}_6+3B^{(1,0)}_6+3B^{(2,0)}_6$\\
     \hline
     $B^{(1)}_9$&$9C^2_1+3C^{(1,0)}_6+3B^{(0,0)}_6+3B^{(1,0)}_6+3B^{(2,0)}_6$&$9C^1_1+3C^{(1,0)}_6+3B^{(0,0)}_6+3B^{(1,0)}_6+3B^{(2,0)}_6$&$9C_1+3C^{(1,0)}_6+3B^{(0,0)}_6+3B^{(1,0)}_6+3B^{(2,0)}_6$\\
     \hline
     $B^{(2)}_9$&$9C^1_1+3C^{(1,0)}_6+3B^{(0,0)}_6+3B^{(1,0)}_6+3B^{(2,0)}_6$&$9C_1+3C^{(1,0)}_6+3B^{(0,0)}_6+3B^{(1,0)}_6+3B^{(2,0)}_6$&$9C^2_1+3C^{(1,0)}_6+3B^{(0,0)}_6+3B^{(1,0)}_6+3B^{(2,0)}_6$\\
     \hline
  \end{tabular}
  }
\end{table}


The $\Delta(54)$ can be realized by higher dimensional fields and string theory similar to $\Delta(27)$.
We introduce twist fields $\sigma_p$ in the same way as $\Delta(27)$.
Then, we consider the reflection, e.g. 
\begin{align}
    \sigma_1 \to \sigma_2, \qquad \sigma_2 \to \sigma_1,
\end{align}
in addition to $\mathbb{Z}_3$ symmetry (\ref{eq:Delta27-1}) and the cyclic symmetry (\ref{eq:Delta27-2}).
The full symmetry becomes $\Delta (54)$ 
\cite{Kobayashi:2006wq,Beye:2014nxa}. 
Hence, the whole symmetry for the multiplication rules of $\Delta(54)$ conjugacy classes is described by $\Delta(54)\rtimes S_4$.

\section{Particle phenomenological aspects}
\label{sec:pheno}

Here, we discuss particle phenomenological aspects of multiplication rules, which were obtained in the previous sections.

\subsection{Coupling selection rules at tree and loop levels}
\label{sec:loop-symmetry}

Here, we study coupling selection rules derived from new multiplication rules.
For comparison, we review briefly conventional coupling selection rules due to group theory.
Suppose that $g_k$ are elements in the group $G$, and they satisfy the following multiplication rule:
\begin{align}
    g_k g_\ell = g_m.
\end{align}
The element $g_m$ appearing in the right hand side is unique.
Each field (particle) $\phi_k$ corresponds to a (representation of) element $g_k$ of $G$.
These fields can couple and the process $\phi_k + \phi_\ell \to \phi_m$ can occur only if $g_k g_\ell = g_m$.
That is charge conservation law. 
These coupling selection rules are not violated in perturbation theory.\footnote{The group symmetry can be broken non-perturbatively if it is anomalous. See e.g. Refs.~\cite{Krauss:1988zc,Ibanez:1991hv,Banks:1991xj,Araki:2008ek,Chen:2015aba,Kobayashi:2021xfs} for anomalies of discrete symmetries.}

We have derived several multiplication rules of classes.
For simplicity, we denote the classes by $C^{(k)}$ and their multiplication rules by 
\begin{align}\label{eq:CC-product}
    C^{(k)}\cdot C^{(\ell)} =\sum_m ~y_{k\ell m}~ C^{(m)},
\end{align}
where $y_{k \ell m}$ are constants and the elements $C^{(m)}$ appearing in the right hand side are not always unique.
Each field (particle) $\phi_k$ corresponds to a class $C^{(k)}$.
Their couplings are allowed and the process $\phi_k + \phi_\ell \to \phi_m$ can occur only if $y_{k \ell m} \neq 0$.
That can lead to new and interesting phenomenological aspects.
An important difference is that the elements $C^{(m)}$ appearing in the right hand side of Eq.~(\ref{eq:CC-product}) are not always unique.

The above new coupling selection rules hold at the tree level.
In general, these coupling selection rules are violated at loop level \cite{Heckman:2024obe,Kaidi:2024wio,Funakoshi:2024uvy}.
Such a phenomenon also leads to another interesting aspect.
That is, some couplings have loop suppression and can be small, if they are forbidden at the tree level.

The perturbation behavior of our selection rules includes more aspects.
For example, the classes of $\mathbb{Z}_2$ gauging of $\mathbb{Z}_N$ have the $\mathbb{Z}_2$ symmetry when $N=$ even.
That is, $C^{(k={\rm even})}$ and $C^{(k={\rm odd})}$ correspond to 
$\mathbb{Z}_2$ even and odd, respectively.
Similarly, the classes of $\mathbb{Z}_3$ gauging of $\mathbb{Z}_N \times \mathbb{Z}'_N$ has the $\mathbb{Z}_3$ symmetry when $N/3=$ integer.
We can find similar symmetries in other multiplication rules.
For these types, we can write multiplication rules by 
\begin{align}\label{eq:CC-product2}
    C^{(k)}_{r_1}\cdot C^{(\ell)}_{r_2} =\sum_m ~y_{k\ell m}~ C^{(m)}_{r_3}.
\end{align}
We have added the new index $r_i$ to denote the $\mathbb{Z}_M$ charges.
Note that $r_3$ in the right hand side is unique because of $\mathbb{Z}_M$ charge conservation law.
This is partially group theoretical selection rule.
Thus, the coupling selection rules due to these $\mathbb{Z}_M$ symmetries are not violated by loop effects.
There are other symmetries in multiplication rules obtained in the previous sections.
For example, there is the permutation symmetry $S_2 \simeq \mathbb{Z}_2$ in $\mathbb{Z}_3$ gauging of $\mathbb{Z}_7$, where $C_1$ is a singlet, and $C_3^1$ and $C_3^2$ are a doublet and they are permuted each other.
This symmetry is not violated by loop effects, either.
For example, we can take the basis to diagonalize this permutation and that is nothing but the above $\mathbb{Z}_2$ even and odd basis.
Obviously, if the theory includes only fields with $C_3^1$ ($C_3^2$), but not $C_3^2$ ($C_3^{(1)}$), there is no $S_2 \simeq \mathbb{Z}_2$ symmetry from the beginning.
We can find similar permutation symmetries in other multiplication rules. 
These symmetries in multiplication rules are shown in Table~\ref{tab:symm-product}. 
Although $S_2$ is isomorphic to $\mathbb{Z}_2$, we use this notation to show the permutation of conjugacy classes, i.e., the outer automorphism. On the other hand, the $\mathbb{Z}_2$ and $\mathbb{Z}_3$ symmetries shown in Table~\ref{tab:symm-product} are represented by diagonal matrices on the classes. 
For example, the multiplication rules in $\mathbb{Z}_3$ gauging of $\mathbb{Z}_3 \times \mathbb{Z}'_3$ has the large symmetry $(\mathbb{Z}_3 \rtimes S_2) \times S_2$.
This symmetry protects some parts of coupling selection rules from violations due to loop effects.
The larger the symmetry is, the larger part is protected from violation due to loop effects.
Obviously, as mentioned above, if the model at the tree level does not have the permutation symmetry like no partner of permutation, such symmetry does not exist from the beginning.
Similarly, we can find the group theoretical symmetries in multiplication rules of conjugacy classes of discrete groups. 
Their conjugacy classes have outer automorphisms and inner automorphisms. 
Their combinations correspond to the full symmetries of multiplication rules.
Those are quite large.
For example, the multiplication rules of $\Delta(27)$ conjugacy classes have the $((\mathbb{Z}_3\times \mathbb{Z}_3) \rtimes Q_8)\rtimes S_3$ symmetry.

\begin{table}[H]
   \centering
   \caption{Symmetries in multiplication rules.}
   \label{tab:symm-product}
   \begin{tabular}{|c|c|}
       \hline
       Classes & Symmetry\\
       \hline
       \hline
       $\mathbb{Z}_2$ gauging of $\mathbb{Z}_{N={\rm even}}$ & $\mathbb{Z}_2$ \\ \hline
       $\mathbb{Z}_3$ gauging of $\mathbb{Z}_7$ & $S_2 $\\
       \hline
       $\mathbb{Z}_3$ gauging of $\mathbb{Z}_3 \times \mathbb{Z}'_3$ & $(\mathbb{Z}_3 \rtimes S_2) \times S_2 $ \\
       \hline
       $\mathbb{Z}_3$ gauging of $\mathbb{Z}_4 \times \mathbb{Z}'_4$ & $S_2\times S_2$ \\
       \hline
       $\mathbb{Z}_3$ gauging $(\mathbb{Z}_2 \times \mathbb{Z}_2' \times \mathbb{Z}_2'')$ & $\mathbb{Z}_2$\\
       \hline
   \end{tabular}
\end{table}

\subsection{$\mathbb{Z}_2$ gauging versus $\mathbb{Z}_3$ gauging}
\label{sec:Z2-Z3}

$\mathbb{Z}_2$ gauging and $\mathbb{Z}_3$ gauging have different properties.
Here, we discuss some differences by using illustrating models.

First we consider the Weinberg operators, which induce the neutrino masses.
We discuss them within the framework of supersymmetric models and write them as 
\begin{align}
    W=c_{ij}L_i H_u L_j H_u,
\end{align}
in the superpotential, 
where $L_i$ denote three generations of lepton doublet superfields and $H_u$ denotes the up-sector Higgs superfield.
For the neutrino masses, supersymmetric and non-supersymmetric models are the same, but later we will discuss the so-called $\mu$-term, that is the supersymmetric mass of $H_u$ and $H_d$, where $H_d$ denotes the down-sector Higgs field.

We briefly review on results in Ref.~\cite{Kobayashi:2025ldi}, 
where selection rules by $\mathbb{Z}_2$ gauging of $\mathbb{Z}_N$ symmetries were studied.
The class $C^{(k)}$ of $\mathbb{Z}_2$ gauging is assigned to each field of $H_u$ and $L_i$ ($i=1,2,3$).
Note that the product $C^{(k)}\cdot C^{(k)}$ always includes $C^{(0)}$.
Furthermore, $C^{(k)}\cdot C^{(k)}\cdot C^{(\ell)}\cdot C^{(\ell)}$ always includes $C^{(0)}$, where $L_i$ may correspond to $C^{(k)}$ and $H_u$ may correspond to $C^{(\ell)}$.
Therefore, the diagonal entries of $c_{ij}$ are always allowed, and the possible textures are \cite{Kobayashi:2025ldi}
\begin{align}\label{eq:teture-Z2}
    \begin{pmatrix}
        * & 0 & 0 \\
        0 & * & 0 \\
        0 & 0 & *
    \end{pmatrix},\quad 
    \begin{pmatrix}
        * & * & 0 \\
        * & * & 0 \\
        0 & 0 & *
    \end{pmatrix},\quad     
        \begin{pmatrix}
        * & 0 & * \\
        0 & * & * \\
        * & * & *
    \end{pmatrix},\quad 
        \begin{pmatrix}
        * & * & * \\
        * & * & * \\
        * & * & *
    \end{pmatrix},\quad 
\end{align}
and their permutations of rows and columns, where the asterisk symbol $*$ denotes non-vanishing elements.

Let us take $\mathbb{Z}_3$ gauging of $\mathbb{Z}_7$ as an example. 
We assign its classes to $L_i$ and $H_u$. 
For example, we assign such that $(L_1,L_2,L_3)$ correspond to $(C_1,C_3^1,C_2^2)$.
When the Higgs field $H_u$ corresponds to $C_1$, 
we obtain
\begin{align}
c_{ij}=
    \begin{pmatrix}
         * & 0 & 0 \\
        0 & 0 & * \\
        0 & * & 0
    \end{pmatrix}.
\end{align}
When  Higgs field $H_u$ corresponds to $C_3^1$, 
we obtain
\begin{align}\label{eq:texture-Z3}
c_{ij}=
    \begin{pmatrix}
         0 & * & * \\
        * & * & * \\
        * & * & *
    \end{pmatrix}.
\end{align}
Thus, $\mathbb{Z}_3$ gauging can lead to mass textures different from ones derived by $\mathbb{Z}_2$ gauging in Ref.~\cite{Kobayashi:2025ldi}.
For example, if we can combine two selection rules, where one corresponds to $\mathbb{Z}_3 $ gauging leading to Eq.~(\ref{eq:texture-Z3}) and $\mathbb{Z}_2$ gauging leading to the third matrix in Eq.~(\ref{eq:teture-Z2}), we can realize the following texture:
\begin{align}\label{eq:texture-Z3-Z2}
c_{ij}=
    \begin{pmatrix}
         0 & 0 & * \\
        0 & * & * \\
        * & * & *
    \end{pmatrix}, 
\end{align}
which corresponds to the neutrino mass in Refs.~\cite{Frampton:2002yf,Fritzsch:2011qv}, i.e. 
the so-called $A_1$ texture.
We will study systematically mass matrices derived from $\mathbb{Z}_3$ gauging elsewhere.

In the supersymmetric standard model, we need the $\mu$-term: 
\begin{align}
    W_\mu = \mu H_u H_d,
\end{align}
in the superpotential.
For example, we assume that $H_u$ corresponds to $C_3^1$.
In this case, if $H_d$ corresponds to $C_3^2$, the $\mu$-term is allowed.
Otherwise, the $\mu$-term is forbidden.
In such a case, we may introduce a singlet field $S$ such that the coupling $\lambda SH_uH_d$ are allowed like the next-to-minimal supersymmetric standard model (NMSSM).
When $H_d$ corresponds to $C_1$, the singlet $S$ must be $C_3^2$.
For this assignment, $S^2$ is forbidden, but $S^3$ is allowed.
That is, the superpotential of the Higgs sector can be written as 
\begin{align}
    W_H=\lambda SH_uH_d + \kappa S^3,
\end{align}
which can realize the Higgs sector in the $\mathbb{Z}_3$ invariant NMSSM.
Note that $\mathbb{Z}_3$ gauging allows the $S^3$ coupling for a singlet $S$ with any class.
When $H_d$ corresponds to $C_3^2$ and the singlet $S$ corresponds to $C_1$, 
the $S^2$ term is also allowed.
This behavior is different from $\mathbb{Z}_2$ gauging.
In $\mathbb{Z}_2$ gauging, the $S^2$ term is always allowed, but the cubic term $S^3$ is not always allowed.

\section{Conclusions}
\label{sec:con}

By clarifying the selection rules of several finite discrete groups, we found that the selection rules of conjugacy classes do not have the conventional group-like structure. When the group is gauged, the gauge invariant classes enjoy finite Abelian or non-Abelian discrete symmetries, as summarized in Table~\ref{tab:symm-product}. The whole conjugacy classes of a discrete group $G$ also enjoy the discrete symmetries determined by the inner and/or outer automorphism of $G$. 
In contrast to the previous studies of $\mathbb{Z}_2$ gauging of $\mathbb{Z}_N$ symmetries, we have systematically analyzed $\mathbb{Z}_3$ gauging of discrete symmetries as well as $S_3$ gauging. 

It turned out that $\mathbb{Z}_3$ gauging can lead to complex representations.
That is different from $\mathbb{Z}_2$ gauging of $\mathbb{Z}_N$ symmetries.
Note that the field $\phi_k$ and its conjugate $\phi^*_k$ correspond to the same class $C^k$ in $\mathbb{Z}_2$ gauging.
The two-point coupling $\phi_k \phi_k$ as well as the $2n$ point self coupling $(\phi_k)^{2n}$ is always allowed.
Also, the diagonal entries in the neutrino mass matrix induced by the Weinberg operators are always allowed \cite{Kobayashi:2025ldi}.
On the other hand, we have the phenomenologically interesting models, where up sector and down sector Higgs fields, $H_u$ and $H_d$ correspond to different classes \cite{Kobayashi:2025znw,Kobayashi:2024cvp}.
$\mathbb{Z}_2$ gauging does not allow the $\mu$-term in supersymmetric standard model when $H_u$ and $H_d$ have classes different from each other.
$\mathbb{Z}_3$ gauging does not always allow the two-point coupling $(\phi_k)^2$ or diagonal entries of neutrino masses by the Weinberg operators, but allow the $3n$ point self coupling $(\phi_k)^{3n}$.
That would lead to interesting neutrino mass textures.
Also, $\mathbb{Z}_3$ gauging may allow the $\mu$-term for proper combinations of classes of $H_u$ and $H_d$, although they are different from each other.
That can lead to a rich structure in mass textures.
Thus, it is quite interesting to apply our $\mathbb{Z}_3$ gauging to particle phenomenology.
We leave it for future work.

\acknowledgments

The authors would like to H.~Okada for useful discussions.
This work was supported by JSPS KAKENHI Grant Numbers JP23K03375 (T.K.) and JP25H01539 (H.O.).

\appendix

\section{$\mathbb{Z}_3$ gauging of $\mathbb{Z}_9$ symmetry}
\label{app:Z3-gauge-9}

Here, we study $\mathbb{Z}_3$ gauging of $\mathbb{Z}_9$.
We denote the generator of $\mathbb{Z}_9$ by $a$, i.e.,
\begin{align}
    a^9=e.
\end{align}
As in Sec.~\ref{sec:Z3gauging-0}, we consider the following $\mathbb{Z}_3$ automorphism $b$:
\begin{align}
    b^{-1}ab=a^7, \qquad b^3=e.
\end{align}
Note that the combination of $N=9$ and $m=7$ satisfies the condition (\ref{eq:condition-Z3}), i.e., $m-1=6=3\times 2$ and $m^2+m+1=57=3 \times 19$.
Then, we define the classes as Eq.~(\ref{eq:class-Z3}), and they satisfy the multiplication rules as Eqs.~(\ref{eq:product-Z3gauging_1}) and (\ref{eq:product-Z3gauging_2}). Note that if $C^{3}_3$ and $C^{6}_3$ appear in the multiplication rules, we have to replace them by $C^{3}_3\Rightarrow 3C^{3}_1,C^{6}_3\Rightarrow 3C^{6}_1$. 
More explicitly, we write the classes as follows:
\begin{align}\label{eq:Z3-Z9class}
    & C_1=\{ e \}, \notag \\
    & C_1^3= \{  a^3 \}, \notag \\
    & C_1^6 = \{ a^6 \}, \notag \\
    & C_3^{(1)}= \{ a, a^4, a^7  \}, \notag \\
    & C_3^{(2)}= \{ a^2, a^5, a^8 \}.
\end{align}
Table~\ref{tab:Z3gaugingZ9} shows their multiplication rules.
That is equivalent to ones of $\mathbb{Z}_3$ of 
$\mathbb{Z}_3 \times \mathbb{Z}'_3$.

  \begin{table}[H]
    \centering
    \caption{Multiplication rules of classes for $\mathbb{Z}_3$ gauging of $\mathbb{Z}_9$.}
    \label{tab:Z3gaugingZ9}
    \begin{tabular}{|c||c|c|c|c|c|}
    \hline
       & $C_1$ & $C^3_1$ & $C^6_1$ & $C^{(1)}_3$ & $C^{(2)}_3$ \\
       \hline
       \hline
       $C_1$& $C_1$ & $C^3_1$ & $C^6_1$ & $C^{(1)}_3$ & $C^{(2)}_3$ \\
       \hline
       $C^3_1$& $C^3_1$& $C^6_1$ & $C_1$ & $C^{(1)}_3$ & $C^{(2)}_3$ \\
       \hline
       $C^6_1$& $C^6_1$ &$C_1$  & $C^3_1$ & $C^{(1)}_3$ & $C^{(2)}_3$ \\
       \hline
       $C^{(1)}_3$& $C^{(1)}_3$ & $C^{(1)}_3$ & $C^{(1)}_3$ & $3C^{(2)}_3$ & $3C_1+3C^3_1+3C^6_1$ \\
       \hline
       $C^{(2)}_3$& $C^{(2)}_3$ & $C^{(2)}_3$ &$C^{(2)}_3$  & $3C_1+3C^3_1+3C^6_1$ & $3C^{(1)}_3$ \\
       \hline  
       \end{tabular}
  \end{table}

The above classes (\ref{eq:Z3-Z9class}) correspond to the conjugacy classes of $\mathbb{Z}_9 \rtimes \mathbb{Z}_3$ without $b$.
The full conjugacy classes of $\mathbb{Z}_9 \rtimes \mathbb{Z}_3$ include additional classes:
\begin{align}
    & B_3^{(1)}=\{ b, ba^3, b a^6\}, \notag \\
    & B_3^{(2)}=\{ ba, ba^4, b a^7\}, \notag \\
    & B_3^{(3)}=\{ ba^2, ba^5, b a^8\}, \notag \\
    & B_3^{(4)}=\{ b^2, b^2a^3, b^2 a^6\}, \notag \\
    & B_3^{(5)}=\{ b^2a, b^2a^4, b^2 a^7\}, \notag \\
    & B_3^{(6)}=\{ b^2a^2, b^2a^5, b^2 a^8\}.  
\end{align}
Their multiplication rules are shown in Table~\ref{tab:Z_9rtimesZ_3}.

\begin{table}[H]
    \centering
    \caption{Multiplication rules of classes for $\mathbb{Z}_9 \rtimes \mathbb{Z}_3$.}
    \label{tab:Z_9rtimesZ_3}
    \resizebox{\textwidth}{!}{
    \begin{tabular}{|c||c|c|c|c|c|c|c|c|c|c|c|}
    \hline
       & $C_1$ & $C^3_1$ & $C^6_1$ & $C^{(1)}_3$ & $C^{(2)}_3$&$B^{(1)}_3$&$B^{(2)}_3$&$B^{(3)}_3$&$B^{(4)}_3$&$B^{(5)}_3$&$B^{(6)}_3$ \\
       \hline
       \hline
       $C_1$& $C_1$ & $C^3_1$ & $C^6_1$ & $C^{(1)}_3$ & $C^{(2)}_3$&$B^{(1)}_3$&$B^{(2)}_3$&$B^{(3)}_3$&$B^{(4)}_3$&$B^{(5)}_3$&$B^{(6)}_3$ \\
       \hline
       $C^3_1$& $C^3_1$& $C^6_1$ & $C_1$ & $C^{(1)}_3$ & $C^{(2)}_3$&$B^{(1)}_3$&$B^{(2)}_3$&$B^{(3)}_3$&$B^{(4)}_3$&$B^{(5)}_3$&$B^{(6)}_3$ \\
       \hline
       $C^6_1$& $C^6_1$ &$C_1$  & $C^3_1$ & $C^{(1)}_3$ & $C^{(2)}_3$&$B^{(1)}_3$&$B^{(2)}_3$&$B^{(3)}_3$&$B^{(4)}_3$&$B^{(5)}_3$&$B^{(6)}_3$ \\
       \hline
       $C^{(1)}_3$& $C^{(1)}_3$ & $C^{(1)}_3$ & $C^{(1)}_3$ & $3C^{(2)}_3$ & $3C_1+3C^3_1+3C^6_1$&$3B^{(2)}_3$&$3B^{(3)}_3$&$3B^{(1)}_3$&$3B^{(5)}_3$&$3B^{(6)}_3$&$3B^{(4)}_3$ \\
       \hline
       $C^{(2)}_3$& $C^{(2)}_3$ & $C^{(2)}_3$ &$C^{(2)}_3$  & $3C_1+3C^3_1+3C^6_1$ & $3C^{(1)}_3$&$3B^{(3)}_3$&$3B^{(1)}_3$&$3B^{(2)}_3$&$3B^{(6)}_3$&$3B^{(4)}_3$&$3B^{(5)}_3$ \\
       \hline       $B^{(1)}_3$&$B^{(1)}_3$&$B^{(1)}_3$&$B^{(1)}_3$&$3B^{(2)}_3$&$3B^{(3)}_3$&$3B^{(4)}_3$&$3B^{(5)}_3$&$3B^{(6)}_3$&$3C_1+3C^3_1+3C^6_1$&$3C^{(1)}_3$&$3C^{(2)}_3$\\
       \hline     $B^{(2)}_3$&$B^{(2)}_3$&$B^{(2)}_3$&$B^{(2)}_3$&$3B^{(3)}_3$&$3B^{(1)}_3$&$3B^{(5)}_3$&$3B^{(6)}_3$&$3B^{(4)}_3$&$3C^{(1)}_3$&$3C^{(2)}_3$&$3C_1+3C^3_1+3C^6_1$\\
       \hline $B^{(3)}_3$&$B^{(3)}_3$&$B^{(3)}_3$&$B^{(3)}_3$&$3B^{(1)}_3$&$3B^{(2)}_3$&$3B^{(6)}_3$&$3B^{(4)}_3$&$3B^{(5)}_3$&$3C^{(2)}_3$&$3C_1+3C^3_1+3C^6_1$&$3C^{(1)}_3$\\
       \hline      $B^{(4)}_3$&$B^{(4)}_3$&$B^{(4)}_3$&$B^{(4)}_3$&$3B^{(5)}_3$&$3B^{(6)}_3$&$3C_1+3C^3_1+3C^6_1$&$3C^{(1)}_3$&$3C^{(2)}_3$&$3B^{(1)}_3$&$3B^{(2)}_3$&$3B^{(3)}_3$\\
       \hline  $B^{(5)}_3$&$B^{(5)}_3$&$B^{(5)}_3$&$B^{(5)}_3$&$3B^{(6)}_3$&$3B^{(4)}_3$&$3C^{(1)}_3$&$3C^{(2)}_3$&$3C_1+3C^3_1+3C^6_1$&$3B^{(2)}_3$&$3B^{(3)}_3$&$3B^{(1)}_3$\\
       \hline    $B^{(6)}_3$&$B^{(6)}_3$&$B^{(6)}_3$&$B^{(6)}_3$&$3B^{(4)}_3$&$3B^{(5)}_3$&$3C^{(2)}_3$&$3C_1+3C^3_1+3C^6_1$&$3C^{(1)}_3$&$3B^{(3)}_3$&$3B^{(1)}_3$&$3B^{(2)}_3$\\
       \hline
       \end{tabular}
       }
  \end{table}

\section{$\Delta(12)\cong A_4$ and $\Delta(24)\cong S_4$}
\label{app:A4S4}

In this section, we list the multiplication rules of $\Delta(12)\cong A_4$ and $\Delta(24)\cong S_4$ conjugacy classes. 
First, $\Delta(12)$ conjugacy classes are chosen as
\begin{align}
    &C_1 = \{ e \},\\
    &C^{(1,0)}_3 = \{ a,a',aa' \},\\
    &C^1_4 = \{ b,ba,ba',baa' \},\\
    &C^2_4 = \{ b^2,b^2a,b^2a',b^2aa' \},
\end{align}
which obey the multiplication rules in Table~\ref{tab:Delta(12)}. 
It turns out that there exists the $S_2\cong \mathbb{Z}_2$ symmetry: $C_4^1 \leftrightarrow C_4^2$, corresponding to the outer automorphism exchanging $b$ and $b^2$.
In addition, there exists $\mathbb{Z}_3$ symmetry: $C_4^1 \rightarrow \omega C_4^1$ and $C_4^2\rightarrow \omega^2 C_4^2$ originating from $b \to \omega b$ with $\omega = e^{2\pi i/3}$. 
Then, the whole symmetry of the multiplication rules is described by $S_3\cong \mathbb{Z}_3\rtimes \mathbb{Z}_2$.

\begin{table}[H]
    \caption{Multiplication rules for conjugacy classes of $\Delta(12)$.}
    \label{tab:Delta(12)}
    \centering
    \begin{tabular}{|c||c|c|c|c|}
    \hline
    &$C_1$&$C^{(1,0)}_3$&$C^1_4$&$C^2_4$\\
    \hline\hline
    $C_1$&$C_1$&$C^{(1,0)}_3$&$C^1_4$&$C^2_4$\\
    \hline
    $C^{(1,0)}_3$&$C^{(1,0)}_3$&$3C_1+2C^{(1,0)}_3$&$3C^1_4$&$3C^2_4$\\
    \hline
    $C^1_4$&$C^1_4$&$3C^1_4$&$4C^2_4$&$4C_1+4C^{(1,0)}_3$\\
    \hline
    $C^2_4$&$C^2_4$&$3C^2_4$&$4C_1+4C^{(1,0)}_3$&$4C^1_4$\\
    \hline
    \end{tabular}
\end{table}

Next, we deal with the $\Delta(24)$ conjugacy classes:
\begin{align}
    &C_1 = \{ e \},\\
    &C^{(1,0)}_3 = \{ a,a',aa' \},\\
    &C_8 = \{ b,ba,ba',baa',b^2,b^2a,b^2a',b^2aa' \},\\
    &B^{(0)}_6 = \{ ca^na'^n,b^2ca^{-n},bca^{-n}|n=0,1 \},\\
    &B^{(1)}_6 = \{ ca^{n+1}a'^n,b^2ca^{-1}a'^{-n-1},bca^{-n}a' |n=0,1\},
\end{align}
which obey the multiplication rules in Table~\ref{tab:Delta(24)}. There exists the $\mathbb{Z}_2\times \mathbb{Z}'_2$ symmetry: $B_6^{(0)} \leftrightarrow B_6^{(1)}$ under $\mathbb{Z}_2$ and $B_6^{(0)}\rightarrow - B_6^{(0)}$, $B_6^{(1)}\rightarrow - B_6^{(1)}$ under $\mathbb{Z}'_2$.
\begin{table}[H]
    \centering
        \caption{Multiplication rules for conjugacy classes of $\Delta(24)$.}
    \label{tab:Delta(24)}
    \resizebox{\textwidth}{!}{
    \begin{tabular}{|c||c|c|c|c|c|}
    \hline
    &$C_1$&$C^{(1,0)}_3$&$C_8$&$B^{(0)}_6$&$B^{(1)}_6$\\
    \hline\hline
    $C_1$&$C_1$&$C^{(1,0)}_3$&$C_8$&$B^{(0)}_6$&$B^{(1)}_6$\\
    \hline
    $C^{(1,0)}_3$&$C^{(1,0)}_3$&$3C_1+2C^{(1,0)}_3$&$C_8$&$B^{(0)}_6+2B^{(1)}_6$&$2B^{(0)}_6+B^{(1)}_6$\\
    \hline
    $C_8$&$C_8$&$C_8$&$8C_1+8C^{(1,0)}_3+4C_8$&$4B^{(0)}_6+4B^{(1)}_6$&$4B^{(0)}_6+4B^{(1)}_6$\\
    \hline
    $B^{(0)}_6$&$B^{(0)}_6$&$B^{(0)}_6+2B^{(1)}_6$&$4B^{(0)}_6+4B^{(1)}_6$&$6C_1+2C^{(1,0)}_3+3C_8$&$4C^{(1,0)}_3+3C_8$\\
    \hline
    $B^{(1)}_6$&$B^{(1)}_6$&$2B^{(0)}_6+B^{(1)}_6$&$4B^{(0)}_6+4B^{(1)}_6$&$4C^{(1,0)}_3+3C_8$&$6C_1+2C^{(1,0)}_3+3C_8$\\
    \hline
    \end{tabular}}
\end{table}

\section{Further discrete gauging}
\label{app:futher}

As summarized in Table~\ref{tab:symm-product}, $\mathbb{Z}_3$ gauging of discrete groups as well as $\mathbb{Z}_2$ has symmetries in multiplication rules.
We can gauge remaining symmetries to obtain other multiplication rules.
For example, in $\mathbb{Z}_3$ gauging of $\mathbb{Z}_3 \times \mathbb{Z}'_3$, we have $S_2$ symmetry, where we exchange $C_3^{(1,0)}$ and $C_3^{(2,0)}$.
When we gauge this $S_2$, the two classes, $C_3^{(1,0)}$ and $C_3^{(2,0)}$ are combined to one, $C_6^{(1,0)}=C_3^{(1,0)} + C_3^{(2,0)}$.
That is nothing but $S_3$ gauging of $\mathbb{Z}_3 \times \mathbb{Z}'_3$.
Similarly, we gauge other cases by remaining symmetries.

Here, we reconsider the above process from $\mathbb{Z}_3$ gauging to $S_3$ gauging.
We choose the basis of classes in $\mathbb{Z}_3$ gauging of $\mathbb{Z}_3 \times \mathbb{Z}'_3$ such that $S_2$ is diagonal, i.e.,
\begin{align}
    C_3^{\pm} = C_3^{(1,0)} \pm C_3^{(2,0)}.
\end{align}
They have $\mathbb{Z}_2$ even and odd charges.
The other classes also have even charges.
The simple $S_2$ gauging is to restrict the representation space to be $\mathbb{Z}_2$ even and project out $\mathbb{Z}_2$ odd, i.e. $C_3^-=0$.
Then, the result is the same as $S_3$ gauging of $\mathbb{Z}_3 \times \mathbb{Z}'_3$ by identifying $C_6^{(1,0)}=C_3^+$.
We may impose a different condition on the represented space such that  $C_3^+=0$.
Then, other classes satisfy the multiplication rules shown in Table~\ref{tab:new}.
Here we normalize $C_3^-$ in a proper way such that all coefficients in multiplication rules are positive, i.e., $C_6'=\frac{i}{\sqrt 6}C_3^-$.
The result is equivalent to the 
$\mathbb{Z}_3$ Tambara-Yamagami fusion rules, i.e. 
\begin{align}
\eta^3=\mathbb{I}, \qquad \eta \cdot {\cal N}= {\cal N}, \qquad {\cal N}^2=\mathbb{I}+\eta + \eta^2,
\end{align}
where we identify $\mathbb{I}=C_1$, $\eta = C_1^1$, $\eta^2 = C_1^2$, and ${\cal N}=C_6'$.
Thus, we have more variety in discrete gauging.

\begin{table}[H]
  \centering
  \caption{New multiplication rules.}
  \label{tab:new}
  \begin{tabular}{|c||c|c|c|c|}
  \hline
     &$C_1$ &$C^1_1$ &$C^2_1$ &$C'_6$ \\
     \hline
     \hline
     $C_1$&$C_1$ &$C^1_1$ &$C^2_1$ &$C'_6$ \\
     \hline
     $C^1_1$& $C^1_1$&$C^2_1$ & $C_1$& $C'_6$\\
     \hline
     $C^2_1$& $C^2_1$&$C_1$ &$C^1_1$ & $C'_6$\\
     \hline
     $C'_6$&$C'_6$ & $C'_6$&$C'_6$ &$C_1+C^1_1+C^2_1$ \\
     \hline
  \end{tabular}
\end{table}

\section{More about $\mathbb{Z}_2$ gauging}
\label{app:Z2}

Here, we study another type of $\mathbb{Z}_2$ gauging of $\mathbb{Z}_N$ symmetries.

Instead of $\mathbb{Z}_2$ gauging of $\mathbb{Z}_N$ reviewed in Sec.~\ref{sec:Z2-gauge-DN}, we use the following automorphism:
\begin{align}
	b_m a^k b_m^{-1}=a^{m-k}.
\end{align}
That is $\mathbb{Z}_2^{(m)}$ gauging.
This is still the $\mathbb{Z}_2$ automorphism but includes a "shift".
The $\mathbb{Z}_2$ gauging in Sec.~\ref{sec:Z2-gauge-DN} with $m=0$ corresponds to geometrical orbifolding around the origin.
On the other hand, $\mathbb{Z}_2^{(m)}$ gauging with $m\neq 0$ corresponds to the geometrical orbifolding at a different point.
Thus we have the conjugacy class:
\begin{align}
	C^{(k)}_m=\{ha^kh^{-1}|h=e,b_m\}=\{a^k,a^{m-k}\}.
\end{align}

Furthermore, we can extend this to $\mathbb{Z}_2^{(a)} \times \mathbb{Z}_2^{(b)}$.
The outer automorphisms are defined as
\begin{align}
	b_p: a^k& \to a^{p-k}, \notag\\
	b_q: a^k&\to a^{q-k},
\end{align}
where the exponents are mod $N$. Note that $b_pb_p=b_qb_q=e$.
We can additionally define the composition of the two actions as
\begin{align}
	b_qb_p=r_\ell: g^k=g^{q-p+k}.
\end{align}
Thus we can construct two types of  elements:
\begin{align}
	b_\ell^n&:a^k=g^{k+n(q-p)} ,\notag \\
	b_pr_\ell^n&:a^k=g^{p-k-n(q-p)}.
\end{align}
Note that when $n<0$, we can get terms like $b_qb_pb_q$.
The class is defined as 
\begin{align}\label{eq:definition-Cpq}
	C^{(k)}_{pq}=\{a^{k+n(q-p)},a^{p-k-n(q-p)}|n\in\mathbb{Z}\}.
\end{align}

Note that the definition of the class in Eq.~(\ref{eq:definition-Cpq}) includes the $\Delta = (p-q)$ modulo structure.
On top of that, recall that $a$ is the generator of $\mathbb{Z}_N$.
Thus, the truly meaningful modulo structure is given by $\Delta'={\rm LCD}(\Delta, N)$, where ${\rm LCD}(k,\ell)$ denotes the least common divisor of $k$ and $\ell$.
For example, when $\Delta$ and $N$ are co-prime each other and $\Delta'=1$, all elements of $Z_N$ group appear in a single class.

The above types of gauging and classes would lead to a new type of selection rules and interesting phenomenology.
We would study it elsewhere.

\bibliography{references}{}
\bibliographystyle{JHEP}

\end{document}